\documentclass{article}
\usepackage{graphicx}
\usepackage{amsmath}
\usepackage{caption}
\usepackage{authblk}
\usepackage[numbers,sort&compress]{natbib}
\usepackage{algorithm}
\usepackage{algpseudocode}
\usepackage{multirow}
\usepackage{datetime}
\usepackage{url}

\title{Encoding Matters: Benchmarking Binary and D-ary Representations for Quantum Combinatorial Optimization}

\author{
Shashank Sanjay Bhat$^{1}$,
Peiyong Wang$^{2}$,
Joseph West$^{1}$ and
Udaya Parampalli$^{1}$
}

\affil{
$^{1}$School of Computing and Information Systems, the University of Melbourne,Parkville VIC 3010, Australia\\
$^{2}$CSIRO Clayton, Research Way, Clayton VIC 3168, Australia
}

\date{\monthname~\the\year}

\begin{document}

\maketitle
\begin{abstract}
    % Combinatorial problems such as the Traveling Salesman Problem (TSP) and the Vehicle Routing Problem (VRP) are commonly formulated via the Quadratic Unconstrained Binary Optimization (QUBO). In this formulation, global constraints are enforced through the penalty terms which require auxiliary variables and increasing the Hamiltonian complexity. We formulate these routing problems using the Quadratic Unconstrained D-ary Optimization (QUDO) where decision variables are represented in higher dimensional Hilbert spaces. We show that QUDO encodes constraints such as route separation, customer uniqueness and depot assignment more naturally than binary formulations. Using a qudit level implementation of the Quantum Approximate Optimization Algorithm, we study TSP and several VRP models. Simulation Experiments are benchmarked against exact classical solutions to demonstrate improved solution quality at a comparable circuit depth, highlighting QUDO as a scalable representation of the routing algorithms. 

    Combinatorial optimization problems are typically formulated using Quadratic Unconstrained Binary Optimization (QUBO), where constraints are enforced through penalty terms that introduce auxiliary variables and rapidly increase Hamiltonian complexity, limiting scalability on near term quantum devices. In this work, we systematically study Quadratic Unconstrained D-ary Optimization (QUDO) as an alternative formulation in which decision variables are encoded directly in higher dimensional Hilbert spaces. We demonstrate that QUDO naturally captures structural constraints across a range of problem classes, including the Traveling Salesman Problem, two variants of the Vehicle Routing Problem, graph coloring, job scheduling, and Max-K-Cut, without the need for extensive penalty constructions. Using a qudit-level implementation of the Quantum Approximate Optimization Algorithm (qudit QAOA), we benchmark these formulations against their binary QUBO counterparts and exact classical solutions. Our study show consistently improved approximation ratios and substantially reduced computational overhead at comparable circuit depths, highlighting QUDO as a scalable and expressive representation for quantum combinatorial optimization.
    
    %of the Traveling Salesman Problem and the Vehicle Routing Problem (VRP) requires a large number of auxiliary variables and penalties, this severely limits the scalability and performance on the NISQ Era devices. We formulate the problem as a Quantum Unconstrained D-ary Optimization (QUDO) and use a simulation based qudit level QAOA to solve the TSP and VRP. QUDO based formulation reduces penalties and offers an advantage in comparison to the QUBO based formulation. 
\end{abstract}

\section{Introduction}
Combinatorial optimization problems such as the Traveling Salesman Problem (TSP) \citep{Laporte1992TSP}, the Vehicle Routing Problem (VRP) \citep{DantzigRamser1959VRP}, graph coloring \citep{graphcolor}, Max-K-Cut \citep{FriezeJerrum1997MaxKCut} and job scheduling \citep{ConwayMaxwellMiller1967Scheduling} are NP-hard in nature and exhibit a rapidly growing solution space as problem size and constraint complexity increase \citep{GareyJohnson1979}. This combinatorial explosion renders many large-scale instances intractable for classical algorithms, motivating sustained interest in quantum approaches for their solution.

Current quantum hardware operates in the Noisy Intermediate-scale quantum (NISQ) regime which is characterized by limited qubit counts, constrained connectivity and errors\citep{Preskill}. Variational algorithms such as Quantum Approximate Optimization Algorithm (QAOA)\citep{Farhi} have emerged as leading candidates for tackling combinatorial problems. The practical performance of QAOA is strongly influenced by how problem constraints are encoded into the underlying Hamiltonian. Standard formulations based on Quadratic Unconstrained Binary Optimization (QUBO)\citep{QUBO} rely on one-hot encodings and penalty terms to enforce global constraints, resulting in large numbers of qubits and densely coupled cost Hamiltonians that challenge near-term implementations. These limitations motivate the exploration of alternative encodings that better align with the native structure of combinatorial problems while remaining compatible with NISQ-era algorithms.

% Existing quantum formulations for the routing problems are based on the binary encoding of the decision variables, typically expressed as a Quadratic Unconstrained Binary Optimization (QUBO) or Ising Hamiltonians. Constraints such as customer uniqueness, depot assignment, vehicle separation are enforced through penalty terms added to the QUBO objective. While this makes the problem structure richer, the resulting Hamiltonian becomes densely coupled and difficult to optimize on the near term quantum hardware. 

% An alternative perspective is to consider how routing decisions are encoded. The Quadratic Unconstrained D-ary Optimization is a generalization of the binary optimization, allowing decision variables to take on integer values which can be represented by qudits.

An alternative perspective is to reconsider how decision variables are encoded in combinatorial optimization problems. Quadratic Unconstrained D-ary Optimization (QUDO) generalizes the binary QUBO framework by allowing decision variables to take values from a finite integer rather than being restricted to binary states \citep{Lucas_2014}. Such variables can be naturally represented using qudits, whose local Hilbert spaces provide direct access to multiple discrete levels \citep{Wang_2020}. This representation enables constraints such as ordering, assignment, and route separation to be expressed intrinsically, reducing the need for auxiliary variables and penalty terms that dominate binary formulations.

In this work, we study the impact of Quadratic Unconstrained D-ary Optimization (QUDO) formulations on both quantum resource requirements and algorithmic performance, in comparison with QUBO encodings. We evaluate these differences using the two metrics: (i) the quantum resources required to represent each problem, and (ii) the performance of the Quantum Approximate Optimization Algorithm (QAOA and its qudit generalization), measured in terms of solution quality and computational cost.
In addition to the approximation ratio relative to the optimal solution, we report the reach percentage which is the fraction of the multi start runs that achieve the target solution quality, the number of optimization iterations and objective function evaluations required to reach the target value, and the total wall clock runtime. Together, these metrics provide a detailed characterization of optimization behavior and scalability across binary and D-ary quantum optimization frameworks.

We begin with the Traveling Salesman Problem, whose QUDO formulation is presented in~\citep{ali2025introductionqudotensorqudo}. For this problem, we provide benchmarking results for both QAOA and qudit QAOA and report the corresponding solution metrics. We then set out to study additional combinatorial optimization problems, including the single-depot vehicle routing problem, multi-depot vehicle routing problem, unweighted Max-$K$-Cut, graph coloring, and job scheduling. For each problem, we present both the QUBO and QUDO formulations and perform a comparative benchmarking analysis using QAOA and qudit QAOA, respectively.

This paper is organized as follows. Section~2 reviews related work and positions the present study within the existing literature. Section~3 describes the methodological framework adopted throughout the paper. Section~4, 5, 6, 7, 8 presents numerical experiments and results for the Traveling Salesman Problem, Two variants of the Vehicle Routing Problem, Max-K-cut, Graph Coloring and Job Scheduling respectively. Finally, Section~9 provides a conclusion and outlines directions for future research.

\section{Related Work}
Mata Ali \citep{ali2025introductionqudotensorqudo} introduced Quadratic Unconstrained D-ary Optimization (QUDO) and its extensions, including Tensor-QUDO and higher-order binary optimization (HOBO), as a unifying modeling framework for discrete and combinatorial optimization problems with categorical or integer-valued variables. The paper carefully develops the formal equivalence between binary, D-ary, and higher-order encodings, and illustrates how problems such as the knapsack problem, the traveling salesman problem, and combinatorial games can be expressed more compactly in a D-ary representation that naturally reflects their underlying structure. The principal contribution lies at the formulation and representation level, highlighting reductions in variable count and constraint overhead relative to traditional QUBO encodings.

Vargas-Calderón \citep{Vargas_Calder_n_2021} proposed a novel mapping of the classical Travelling Salesman Problem (TSP) onto a many-qudit Hamiltonian, showing that an $N$-city TSP can be encoded in an $N^N$-dimensional qudit Hilbert space rather than the $2^{N^2}$ space arising from standard QUBO one-hot encodings. They argue that this exponential reduction in Hilbert space size may facilitate both quantum and classical solution methods. The authors explicitly construct a qudit Hamiltonian whose ground state corresponds to a valid TSP tour and validate the approach using classical simulations based on variational Monte Carlo with neural quantum states, solving instances with up to $\sim 100$ cities. These results demonstrate that the many-qudit representation can yield computational advantages in a classical simulation context.

Deller \citep{Deller_2023} developed a generalization of the Quantum Approximate Optimization Algorithm (QAOA) to qudit systems and show how integer optimization problems can be encoded directly in such multi-level quantum systems. They describe the extended QAOA framework on a Hilbert space of dimension $d^N$, including cost Hamiltonians constructed using generalized Pauli and angular-momentum operators to represent multi-valued cost functions, and discuss three strategies for incorporating linear constraints into the variational circuit, namely penalty terms, ancilla-assisted conditionals, and dynamical decoupling. As a proof of concept, the authors present numerical simulations for a simplified electric-vehicle charging optimization problem, mapped to a max-$k$ graph coloring instance, illustrating the flexibility of qudit formulations for a range of integer optimization problems.

Weggemans et al \citep{Weggemans_2022} introduced a qudit-native formulation of correlation clustering and demonstrate how it can be solved with a qudit-based QAOA implemented on a neutral-atom (Rydberg) platform. They map each clustering decision to a multi-level (qudit) degree of freedom, thereby avoiding the overhead of one-hot or binary encodings, and construct both cost and mixer Hamiltonians that naturally operate within the feasible subspace. The authors develop a full-stack pipeline spanning from high-level problem encoding to low-level gate decomposition tailored to Rydberg qudit hardware, and provide resource counts and circuit depth estimates that illustrate the potential hardware efficiency of qudit implementations. Through this concrete case study, the work highlights practical advantages of qudit representations, including reduced gate counts and more compact formulations, when tackling non-binary combinatorial problems.

Xavier et al.\citep{xavier2025a} investigated how Quadratic Integer Programs (QIPs) can be reformulated using either Quadratic Unconstrained Integer Optimization (QUIO) or traditional Quadratic Unconstrained Binary Optimization (QUBO), emphasizing how integer decision variables can be represented natively for qudit-compatible hardware versus binary qubit-based systems. They define several representative QIP classes such as the Quadratic Facility Location, Quadratic Inventory Management, Quadratic Vehicle Routing, and Quadratic Knapsack problems. The paper empirically compares the resulting QUIO and QUBO reformulations in terms of variable count, problem density, and performance when solved on both classical solvers and prototype qudit devices such as the Entropy Dirac-3 quantum computer.

Prior work on D-ary encodings and qudit-based optimization highlights important representational and hardware-level advantages, but remains limited either to formulation level analyses without algorithmic evaluation, to qudit-QAOA demonstrations on isolated problem instances, or to empirical comparisons focused on variable counts and resource metrics rather than solution quality. Existing studies do not provide a systematic, head-to-head assessment of QUDO versus QUBO within a unified variational framework, nor do they quantify how representation choice affects approximation ratio, feasibility, and scalability of QAOA across multiple canonical NP-hard combinatorial problems under comparable circuit depths. In contrast, the present work bridges these gaps by combining explicit QUDO and QUBO formulations with qudit and qubit based QAOA implementations across a broad problem suite, enabling a controlled evaluation of the algorithmic consequences of d-ary encodings and demonstrating that QUDO paired with qudit-QAOA can consistently yield superior approximation performance and robustness relative to standard binary approaches.

\section{Methodology }

In this work, we adopt a unified framework to compare binary and D-ary quantum formulations across a range of combinatorial optimization problems. For each problem instance, we construct both QUBO and QUDO formulations. We explicitly describe how decision variables and constraints are encoded in each case. We then analyze and compare the corresponding quantum resource requirements. Particular emphasis is placed on how the dimension of the accessible Hilbert space scales with the number of qubits in the QUBO formulation and the number of qudits in the QUDO formulation.

To evaluate algorithmic performance, we implement QAOA for QUBO encodings and qudit QAOA for QUDO encodings. All simulations are performed at circuit depths $p=1,2,3$, with variational parameters ${\gamma_\ell,\beta_\ell}_{\ell=1}^p$ optimized using the COBYLA classical optimizer~\citep{Powell1994COBYLA}. Since the QAOA optimization landscape is highly nonconvex, certain parameter choices may not yield high quality solutions \citep{QAOA_Zhou}. To mitigate sensitivity to parameter initialization, reduce bias from suboptimal local minima, and limit computational overhead, we perform ten independent multi-start optimization runs with randomly initialized parameters.

To mitigate sensitivity to parameter initialization, reduce bias from suboptimal local minima, and limit computational overhead, we perform ten independent multi-start optimization runs with randomly initialized parameters.

Qubit QAOA and qudit QAOA share the same variational structure. The difference lies in how decision variables are encoded. QUBO formulations typically require a larger number of penalty terms to enforce constraints. QUDO formulations exploit the larger d-level Hilbert space. This allows constraints to be encoded more directly. As a result, fewer penalty terms are often required.

In qubit QAOA, the cost Hamiltonian is implemented using two qubit entangling gates such as CNOT or CZ. These gates are combined with single qubit RZ rotations. The mixer Hamiltonian is implemented using single qubit RX rotations. In the qudit generalization, the gates are replaced by their higher dimensional counterparts. The initial uniform superposition is prepared using a K-dimensional discrete Fourier transform. The cost layer is implemented as a two qudit diagonal operator. The mixer layer is generated by a cyclic shift Hamiltonian acting on d-level systems. This mixer is implemented as a dense single qudit unitary. The qudit QAOA circuits are implemented using Cirq~\citep{CirqDevelopers2025}. The fourier\_gate function prepares the initial superposition. The qudit\_mixer\_gate function implements the mixer layer and the equal\_label\_edge\_gate function encodes the cost Hamiltonian.

Performance is evaluated using multiple metrics. We report the approximation ratio relative to the exact optimal solution when available. We also record the number of optimizer iterations and function evaluations required to reach a specified tolerance. This provides a measure of optimization efficiency. We report the reach percentage. This is defined as the fraction of multi-start runs that achieve the target solution quality. We also report the feasibility probability $P_{\mathrm{valid}}$. This corresponds to the probability of sampling a valid solution from the final quantum state. Finally, we measure the total wall-clock time required for parameter optimization and circuit evaluation. This provides a practical assessment of computational scalability.

We consider several combinatorial optimization problems in our experiments. The optimization problems considered are the Traveling Salesman Problem (TSP), single depot Vehicle Routing Problem (VRP), multi depot VRP, Max-K-Cut, Graph coloring and Job scheduling. All of the experiments are performed in a noiseless environment.The following section presents the detailed results for the Traveling Salesman Problem.

\section{Traveling Salesman Problem}
% The Traveling Salesman problem is a combinatorial optimization problem in which a single agent must visit a set of cities exactly once and return back to the starting point while minimizing the total travel distance\cite{Laporte1992TSP}. Let $N$ denote the number of cities and $D_{i,k}$ the distance between cities $i$ and $k$.
% Indices are taken modulo $N$.

The Traveling Salesman Problem is a canonical combinatorial optimization problem in which a single agent must visit a set of $N$ cities exactly once and return to the starting city while minimizing the total travel distance \citep{Laporte1992TSP}. Let $D_{i,k}$ denote the distance between cities $i$ and $k$. Throughout this work, tour indices are taken modulo $N$ to represent the cyclic nature of the tour. Table~\ref{tab:tsp_formulation} summarizes the Hamiltonian formulations of the TSP under QUBO and QUDO encodings, while a detailed derivation of each Hamiltonian is provided in the Appendix A.

\begin{table}[H]
\centering
\caption{QUBO and QUDO formulations of TSP.}
\label{tab:tsp_formulation}

\small
\begin{tabular}{p{0.47\columnwidth} | p{0.47\columnwidth}}
\hline
\textbf{QUBO formulation} & \textbf{QUDO formulation} \\
\hline

\textbf{Decision variables} &
\textbf{Decision variables} \\[3pt]

$X_{i,j} \in \{0,1\}$ indicates whether city $i$ is visited at tour position $j$.
A valid tour is enforced via permutation penalties. &
$v_j \in \{1,\dots,N\}$ directly denotes the city visited at tour position $j$.
Each variable is a $d$-ary qudit with $d=N$. \\[8pt]

\textbf{Hamiltonian} &
\textbf{Hamiltonian} \\[3pt]

\[
\begin{aligned}
H^{\mathrm{QUBO}} =
&\sum_{j=1}^{N}
\sum_{i,k=1}^{N}
D_{i,k}\,
X_{i,j} X_{k,(j+1)\bmod N} \\
&+ A \sum_{j=1}^{N}
\Big(\sum_{i=1}^{N} X_{i,j}-1\Big)^2 \\
&+ A \sum_{i=1}^{N}
\Big(\sum_{j=1}^{N} X_{i,j}-1\Big)^2
\end{aligned}
\]

\emph{The cost term encodes the tour length. Penalty terms enforce
one city per position and one visit per city.}
\citep{Lucas_2014,SmithMiles2025TSPQuantum}
&
\[
\begin{aligned}
H^{\mathrm{QUDO}} =
&\sum_{j=1}^{N}
D_{v_j,\,v_{(j+1)\bmod N}} \\
&+ A \sum_{1 \le p < q \le N}
\delta(v_p = v_q)
\end{aligned}
\]

\emph{Tour length is evaluated directly. Collision penalties suppress
repeated city assignments without one-hot constraints.}
\\

\hline
\end{tabular}
\end{table}

\subsection{Resource requirements}
Table~\ref{tab:qubo_qudo_hilbert} compares the quantum resource requirements of the QUBO and QUDO formulations of the Traveling Salesman Problem. In the QUBO encoding, $N^2$ binary variables are required to represent the permutation matrix, resulting in a Hilbert space that scales as $2^{N^2}$. In contrast, the QUDO formulation employs only $N$ $d$-level variables with $d=N$, yielding a Hilbert space of size $N^N$ \citep{Vargas_Calder_n_2021}. Although both spaces grow exponentially, the QUDO encoding achieves a substantial reduction in variable count and constraint overhead, leading to a significantly more compact Hamiltonian representation and improved suitability for near-term quantum algorithms.

\begin{table}[H]
\centering
\caption{Quantum resources and Hilbert space sizes for QUBO and QUDO formulations of the Traveling Salesman Problem.
The Hilbert space dimension corresponds to the total number of computational basis states.
QUBO uses $N^2$ qubits, while QUDO uses $N$ qudits of local dimension $d=N$.}
\label{tab:qubo_qudo_hilbert}
\resizebox{\columnwidth}{!}{
\begin{tabular}{c|c|c|c|c}
\hline
$N$
& \begin{tabular}[c]{@{}c@{}}QUBO qubits\\[-2pt]$(N^2)$\end{tabular}
& \begin{tabular}[c]{@{}c@{}}QUBO Hilbert space\\[-2pt]$(2^{N^2})$\end{tabular}
& \begin{tabular}[c]{@{}c@{}}QUDO qudits\\[-2pt]$(N)$\end{tabular}
& \begin{tabular}[c]{@{}c@{}}QUDO Hilbert space\\[-2pt]$(N^{N})$\end{tabular} \\
\hline
3 & 9  & $2^{9}=512$                     & 3 & $3^{3}=27$ \\
4 & 16 & $2^{16}=65{,}536$               & 4 & $4^{4}=256$ \\
5 & 25 & $2^{25}\approx3.3\times10^{7}$  & 5 & $5^{5}=3{,}125$ \\
6 & 36 & $2^{36}\approx6.9\times10^{10}$ & 6 & $6^{6}=46{,}656$ \\
7 & 49 & $2^{49}\approx5.6\times10^{14}$ & 7 & $7^{7}=823{,}543$ \\
\hline
\end{tabular}
}
\end{table}
% \begin{table}[H]
% \centering
% \caption{Quantum resources and Hilbert space sizes for QUBO and QUDO formulations of the TSP.
% The Hilbert space dimension corresponds to the total number of computational basis states.
% QUBO requires $N^2$ qubits, while QUDO uses $N$ qudits of dimension $d=N$.}
% \label{tab:qubo_qudo_hilbert}
% \resizebox{\columnwidth}{!}{
% \begin{tabular}{c|cc|ccc}
% \hline
%  & \multicolumn{2}{c|}{QUBO} & \multicolumn{3}{c}{QUDO} \\
% $N$
% & \# qubits $(N^2)$
% & Hilbert space $(2^{N^2})$
% & \# qudits $(N)$
% & qudit dim. $(d=N)$
% & Hilbert space $(N^{N})$ \\
% \hline
% 3 & $3^2=9$  
%   & $2^{9}=512$
%   & $3$
%   & $3$
%   & $3^{3}=27$ \\

% 4 & $4^2=16$
%   & $2^{16}=65{,}536$
%   & $4$
%   & $4$
%   & $4^{4}=256$ \\

% 5 & $5^2=25$
%   & $2^{25}\approx3.3\times10^{7}$
%   & $5$
%   & $5$
%   & $5^{5}=3{,}125$ \\

% 6 & $6^2=36$
%   & $2^{36}\approx6.9\times10^{10}$
%   & $6$
%   & $6$
%   & $6^{6}=46{,}656$ \\

% 7 & $7^2=49$
%   & $2^{49}\approx5.6\times10^{14}$
%   & $7$
%   & $7$
%   & $7^{7}=823{,}543$ \\
% \hline
% \end{tabular}
% }
% \end{table}

\subsection{Experiment with QAOA and qudit QAOA}

\begin{table}[H]
\centering
\caption{Performance summary of QUBO and QAOA for different circuit depths $p$.
Reported values are mean $\pm$ standard deviation over 10 random starts.}
\label{tab:qubo_qaoa_summary}
\resizebox{\columnwidth}{!}{
\begin{tabular}{c c c c c c c c}
\hline
$p$ & $N$ 
& AR 
& Reach (\%) 
& Steps to target 
& Evals to target 
& $P_{\text{valid}}$ 
& Time (s) \\
\hline
\multirow{3}{*}{1}
& 3 & $1.0000\pm0.0000$ & 100 & $1.0\pm0.0$ & $1.0\pm0.0$ & $0.0386\pm0.0293$ & $0.40\pm0.08$ \\
& 4 & $1.0378\pm0.0935$ & 60  & $8.8\pm5.2$ & $7.8\pm4.8$ & $0.0006\pm0.0005$ & $1.50\pm0.29$ \\
& 5 & --               & 0   & --           & --           & $0.0000\pm0.0000$ & $219.95\pm18.33$ \\
\hline
\multirow{3}{*}{2}
& 3 & $1.0000\pm0.0000$ & 100 & $1.0\pm0.0$ & $1.0\pm0.0$ & $0.0343\pm0.0268$ & $0.51\pm0.02$ \\
& 4 & $1.0031\pm0.0093$ & 80  & $9.0\pm5.8$ & $7.8\pm5.6$ & $0.0009\pm0.0007$ & $2.47\pm0.25$ \\
& 5 & --               & 0   & --           & --           & $0.0000\pm0.0000$ & $745.93\pm47.13$ \\
\hline
\multirow{3}{*}{3}
& 3 & $1.0000\pm0.0000$ & 100 & $1.2\pm0.6$ & $1.1\pm0.3$ & $0.0291\pm0.0150$ & $0.63\pm0.01$ \\
& 4 & $1.0083\pm0.0134$ & 70  & $12.4\pm8.8$ & $10.3\pm7.3$ & $0.0006\pm0.0003$ & $3.06\pm0.15$ \\
& 5 & --               & 0   & --           & --           & $0.0000\pm0.0000$ & $1544.11\pm76.15$ \\
\hline
\end{tabular}
}
\end{table}

\begin{table}[H]
\centering
\caption{Performance summary of QUDO and qudit QAOA for different circuit depths $p$.
Reported values are mean $\pm$ standard deviation over 10 random starts.}
\label{tab:qudo_dqaoa_summary}
\resizebox{\columnwidth}{!}{
\begin{tabular}{c c c c c c c c}
\hline
$p$ & $N$ 
& AR  
& Reach (\%) 
& Steps to target 
& Evals to target 
& $P_{\text{valid}}$ 
& Time (s) \\
\hline
\multirow{5}{*}{1}
& 3 & $1.0000\pm0.0000$ & 100 & $1.2\pm0.6$ & $1.1\pm0.3$ & $0.1767\pm0.0675$ & $0.10\pm0.09$ \\
& 4 & $1.0000\pm0.0000$ & 80  & $1.2\pm0.7$ & $1.1\pm0.4$ & $0.0510\pm0.0415$ & $0.08\pm0.01$ \\
& 5 & $1.0051\pm0.0094$ & 60  & $3.5\pm4.5$ & $2.7\pm2.9$ & $0.0151\pm0.0147$ & $0.19\pm0.03$ \\
& 6 & $1.0905\pm0.0833$ & 20  & $7.0\pm1.4$ & $6.0\pm2.8$ & $0.0025\pm0.0024$ & $2.34\pm0.79$ \\
& 7 & $1.2169\pm0.0652$ & 0   & --           & --           & $0.0001\pm0.0003$ & $41.57\pm15.93$ \\
\hline
\multirow{5}{*}{2}
& 3 & $1.0000\pm0.0000$ & 100 & $1.0\pm0.0$ & $1.0\pm0.0$ & $0.1511\pm0.0792$ & $0.14\pm0.03$ \\
& 4 & $1.0000\pm0.0000$ & 90  & $1.4\pm1.3$ & $1.2\pm0.7$ & $0.0543\pm0.0339$ & $0.17\pm0.02$ \\
& 5 & $1.0014\pm0.0044$ & 90  & $3.9\pm3.0$ & $3.6\pm2.7$ & $0.0159\pm0.0070$ & $0.67\pm0.07$ \\
& 6 & $1.0157\pm0.0250$ & 30  & $32.7\pm5.9$ & $30.3\pm4.7$ & $0.0054\pm0.0042$ & $8.87\pm0.49$ \\
& 7 & $1.1015\pm0.0750$ & 0   & --           & --           & $0.0044\pm0.0036$ & $235.43\pm43.79$ \\
\hline
\multirow{5}{*}{3}
& 3 & $1.0000\pm0.0000$ & 100 & $1.0\pm0.0$ & $1.0\pm0.0$ & $0.1637\pm0.0765$ & $0.24\pm0.03$ \\
& 4 & $1.0000\pm0.0000$ & 100 & $1.2\pm0.6$ & $1.2\pm0.6$ & $0.0507\pm0.0274$ & $0.32\pm0.02$ \\
& 5 & $1.0000\pm0.0000$ & 100 & $6.0\pm4.2$ & $4.7\pm3.0$ & $0.0215\pm0.0111$ & $0.97\pm0.06$ \\
& 6 & $1.0026\pm0.0084$ & 80  & $16.8\pm13.5$ & $14.8\pm12.3$ & $0.0079\pm0.0049$ & $13.65\pm0.85$ \\
& 7 & $1.0770\pm0.0564$ & 10  & $11.0\pm0.0$ & $9.0\pm0.0$ & $0.0070\pm0.0061$ & $368.00\pm25.93$ \\
\hline
\end{tabular}
}
\end{table}
To assess the impact of binary versus d-ary encodings, we compare standard QAOA applied to the QUBO formulation with qudit QAOA applied to the QUDO formulation of TSP. Approximation ratios, feasibility probabilities, optimization effort, and wall-clock runtimes for circuit depths p = 1, 2, 3 are reported in Tables~\ref{tab:qubo_qaoa_summary} and~\ref{tab:qudo_dqaoa_summary}.

For small problem sizes, both formulations reliably converge to optimal or near-optimal solutions, typically within a small number of optimization steps and circuit evaluations. As the number of cities increases, the two approaches exhibit different optimization behavior, indicating differences in the quality of their variational landscapes.

In the QUBO based formulation, the number of optimization steps and objective function evaluations required to reach a target solution increases rapidly with problem size and circuit depth. This increase is accompanied by a reduction in the probability of sampling valid tours. As a result, the overall optimization process becomes increasingly costly, leading to growing wall clock runtimes and rendering simulations impractical for larger instances at higher depths.

In contrast, the QUDO based qudit QAOA consistently requires fewer optimization steps and evaluations to reach the target, even as the problem size grows. The higher probability of sampling feasible solutions throughout the optimization indicates that a larger fraction of the parameter space yields informative objective evaluations. Taken together, these observations suggest that the variational landscape induced by the d-ary encoding is of higher effective quality for classical optimization, enabling more efficient parameter updates and faster convergence. Increasing the circuit depth in the qudit QAOA setting leads to improved approximation ratios without a corresponding explosion in optimization cost.

In some instances, both QUBO and QUDO exhibit undefined steps to target and evaluations to target metrics because the optimizer does not reach the exact optimum within the allotted computational budget. However, a well-defined approximation ratio can still be reported, as it is computed from the solution value attained during the optimization rather than from successful convergence to the optimum. 

In the next section, we extend this analysis to the Vehicle Routing Problem, which generalizes the TSP by requiring the simultaneous optimization of multiple routes over a shared set of customers.

\section{Vehicle Routing Problem}
The Vehicle Routing Problem extends the Traveling Salesman Problem by introducing multiple vehicles that must jointly serve a set of customers. In contrast to TSP where a single tour gets optimized, VRP requires determining the assignment of the customers to vehicles and ordering of the visits per route. Even in a single depot scenario, the coupling between the routing and partitioning increases the combinatorial complexity \citep{DantzigRamser1959VRP,Toth2014VRP}. As a result, formulating the VRP within a quadratic unconstrained framework introduces substantially more decision variables and penalty terms than in the TSP. In the following sections we study variants of VRP which involves a single depot and multi depot with no capacity or time based constraints. 

\subsection{Single Depot VRP}
We consider the single depot vehicle routing problem (VRP) with N customers and V identical vehicles. All vehicles start and end at a common depot labeled 0. The objective is to determine a set of V routes that collectively visit each customer exactly once while minimizing the total travel cost \citep{larson_odoni_urban_operations_research_1981}. Table \ref{tab:vrp_formulation}
summarizes the Hamiltonian formulations of the single depot VRP under QUBO and QUDO encodings. A detailed derivation of each Hamiltonian is provided in the Appendix B.

\begin{table}[H]
\centering
\caption{QUBO and QUDO formulations of the single-depot Vehicle Routing Problem (VRP).
Here $N$ denotes the number of customers and $V$ the number of vehicles.}
\label{tab:vrp_formulation}
\resizebox{\columnwidth}{!}{
\begin{tabular}{p{0.47\columnwidth}|p{0.47\columnwidth}}
\hline
\textbf{QUBO formulation} & \textbf{QUDO formulation} \\
\hline

\textbf{Decision variables} &
\textbf{Decision variables} \\[6pt]

Binary variables
\[
\begin{aligned}
X_{i,j} &\in \{0,1\} \\
i &\in \{0,\dots,N\} \\
j &\in \{1,\dots,M\}
\end{aligned}
\]
where $X_{i,j}=1$ indicates node $i$ occupies position $j$.
&
Discrete variables
\[
v_j\in\{0,\dots,N\},\quad
j\in\{1,\dots,M\},
\]
where $v_j$ stores the node label at position $j$. \\[10pt]

\textbf{Hamiltonian} &
\textbf{Hamiltonian} \\[6pt]

\[
\begin{aligned}
H_{\mathrm{VRP}}^{\mathrm{QUBO}} =\;&
\sum_{j,i,k} D_{i,k}\, X_{i,j} X_{k,j+1} \\
&+ A \sum_j \Big(\sum_i X_{i,j}-1\Big)^2 \\
&+ A \sum_i \Big(\sum_j X_{i,j}-1\Big)^2 \\
&+ B \Big(\sum_j X_{0,j}-V\Big)^2
\end{aligned}
\]
&
\[
\begin{aligned}
H_{\mathrm{VRP}}^{\mathrm{QUDO}} =\;&
\sum_j D_{v_j,v_{j+1}} \\
&+ A \sum_{i=1}^{N}\sum_{p<q}
\delta(v_p=i)\,\delta(v_q=i) \\
&+ B \Big(\sum_j \delta(v_j=0)-V\Big)^2
\end{aligned}
\] \\[12pt]

\textbf{Constraint encoding} &
\textbf{Constraint encoding} \\[6pt]

Constraints enforced via one-hot penalty terms:
\begin{itemize}
\item one node per position
\item each customer visited once
\item depot appears $V$ times
\end{itemize}
&
Constraints embedded in the variable domain:
\begin{itemize}
\item exactly one node per position by construction
\item customer uniqueness via collision penalties
\item depot occurrences encode route separation
\end{itemize} \\

\hline
\end{tabular}
}
\end{table}

\subsubsection{Resource requirements}

Table \ref{tab:vrp_resources} shows the resource requirements of the single depot VRP between the QUBO and QUDO formulations. In the QUBO approach, the routing sequence is encoded using binary variables $X_{i,j}\in\{0,1\}$, where $i\in\{0,\dots,N\}$ labels the $N$ customers and the depot, and $j\in\{1,\dots,N+V\}$ labels positions in the sequence. This one-hot encoding requires $(N+1)(N+V)$ qubits, since each position must store a binary indicator for every possible node. As a result, the size of the Hilbert space scales as $2^{(N+1)(N+V)}$, and the associated Hamiltonian contains dense quadratic couplings arising from permutation and depot-count constraints.

The QUDO formulation replaces the one-hot encoding with discrete-valued variables $v_j\in\{0,\dots,N\}$, so that each position in the sequence is represented by a single qudit of local dimension $d=N+1$. This reduces the number of quantum variables to $N+V$ qudits, with a total Hilbert space dimension of $(N+1)^{N+V}$. While this introduces higher-dimensional local degrees of freedom, it eliminates an extensive number of auxiliary binary variables and constraint penalties. The QUDO formulation achieves a reduction in the resources required and yields a more compact representation that is particularly advantageous for near term qudit based quantum processors.

\begin{table}[h]
\centering
\caption{Quantum resource requirements for QUBO and QUDO formulations of the single-depot VRP.
Here $N$ denotes the number of customers, $V$ the number of vehicles, and the sequence length is
$M=N+V$.}
\label{tab:vrp_resources}
\resizebox{\columnwidth}{!}{
\begin{tabular}{cc|c|c|c|c}
\hline
$N$ & $V$
& \begin{tabular}[c]{@{}c@{}}QUBO qubits\\[-2pt]$((N{+}1)(N{+}V))$\end{tabular}
& \begin{tabular}[c]{@{}c@{}}QUBO Hilbert space\\[-2pt]$(2^{(N+1)(N+V)})$\end{tabular}
& \begin{tabular}[c]{@{}c@{}}QUDO qudits\\[-2pt]$(N{+}V)$\end{tabular}
& \begin{tabular}[c]{@{}c@{}}QUDO Hilbert space\\[-2pt]$((N{+}1)^{N+V})$\end{tabular} \\
\hline
3 & 2 & 20 & $2^{20}\!\approx\!1.05\times10^{6}$  & 5 & $4^{5}=1.02\times10^{3}$ \\
4 & 2 & 30 & $2^{30}\!\approx\!1.07\times10^{9}$  & 6 & $5^{6}=1.56\times10^{4}$ \\
4 & 3 & 35 & $2^{35}\!\approx\!3.44\times10^{10}$ & 7 & $5^{7}=7.81\times10^{4}$ \\
5 & 3 & 48 & $2^{48}\!\approx\!2.81\times10^{14}$ & 8 & $6^{8}=1.68\times10^{6}$ \\
\hline
\end{tabular}
}
\end{table}
% \begin{table}[h]
% \centering
% \caption{Quantum resource requirements for QUBO and QUDO formulations of the single-depot VRP.
% Here $N$ denotes the number of customers and $V$ the number of vehicles, with sequence length
% $M=N+V$.}
% \label{tab:vrp_resources}
% \resizebox{\columnwidth}{!}{
% \begin{tabular}{cc|cc|cc}
% \hline
% $N$ & $V$ 
% & QUBO qubits $(N{+}1)(N{+}V)$
% & QUBO Hilbert space $2^{(N+1)(N+V)}$
% & QUDO qudits $(N{+}V)$
% & QUDO Hilbert space $(N{+}1)^{N+V}$ \\
% \hline
% 3 & 2 
% & $(3{+}1)(3{+}2)=20$
% & $2^{20}\!\approx\!1.05\times10^{6}$
% & $3{+}2=5$
% & $4^{5}=1.02\times10^{3}$ \\

% 4 & 2 
% & $(4{+}1)(4{+}2)=30$
% & $2^{30}\!\approx\!1.07\times10^{9}$
% & $4{+}2=6$
% & $5^{6}=1.56\times10^{4}$ \\

% 4 & 3 
% & $(4{+}1)(4{+}3)=35$
% & $2^{35}\!\approx\!3.44\times10^{10}$
% & $4{+}3=7$
% & $5^{7}=7.81\times10^{4}$ \\

% 5 & 3 
% & $(5{+}1)(5{+}3)=48$
% & $2^{48}\!\approx\!2.81\times10^{14}$
% & $5{+}3=8$
% & $6^{8}=1.68\times10^{6}$ \\
% \hline
% \end{tabular}
% }
% \end{table}

\subsubsection{Experiment with QAOA and qudit QAOA}
Tables \ref{tab:qubo_qaoa_svrp_summary} and \ref{tab:qudo_dqaoa_svrp_summary} compare the performance of the QUDO formulation solved using qudit QAOA with that of the standard QUBO formulation solved using qubit based QAOA for the single depot VRP. Across all tested instances and circuit depths, the QUDO formulation consistently produces feasible solutions with approximation ratios close to unity. Increasing the circuit depth systematically improves solution quality, with the exact optimum achieved for all tested instances at $p = 2$ and $p = 3$.

The QUBO-based formulation exhibits extremely low probabilities of sampling feasible solutions, even for the smallest problem sizes considered. In most cases, the optimizer fails to encounter a valid solution within the allotted optimization budget. As a result, the metrics "steps to target" and "evaluations to target" are undefined and reported as "-" but approximation ratios are reported in some of these cases. These values correspond to the best objective value observed during the optimization process, irrespective of feasibility. The presence of "-" entried for steps and evaluations and a defined approximate ratio highlight the difficulty of the QUBO formulation.

While the single depot VRP already demonstrates the advantages of the QUDO formulation, many real-world routing applications involve additional structural complexity. In particular, multi depot scenarios, selection of departure points, and construction of valid routes. In the following subsection, we extend the analysis to the multi depot vehicle routing problem and examine how the QUBO and QUDO formulations scale under this increased combinatorial complexity.

\begin{table}[H]
\centering
\caption{Performance summary of QUBO and QAOA for the single depot VRP. Reported values are mean $\pm$ standard deviation over 10 random starts.}
\label{tab:qubo_qaoa_svrp_summary}
\resizebox{\columnwidth}{!}{
\large
\begin{tabular}{c c c c c c c c c c}
\hline
$p$ & $N$ & $V$
& AR 
& Reach (\%) 
& Steps to target 
& Evals to target 
& $P_{\text{valid}}$ 
& Time (s) 
& Best cost \\
\hline
1 & 3 & 2 & $1.5037\pm0.0916$ & 0 & $27.0\pm0.0$ & $27.0\pm0.0$ & $0.0001\pm0.0002$ & $18.47\pm1.58$ & 45.5000 \\
1 & 3 & 3 & -                & 0 & -           & -           & $0.0000\pm0.0000$ & $320.53\pm32.01$ & - \\
\hline
2 & 3 & 2 & $1.2910\pm0.2015$ & 0 & -           & -           & $0.0001\pm0.0002$ & $33.90\pm0.91$ & 36.5000 \\
2 & 3 & 3 & -                & 0 & -           & -           & $0.0000\pm0.0000$ & $672.86\pm11.16$ & - \\
\hline
3 & 3 & 2 & $1.3493\pm0.2146$ & 0 & $14.0\pm8.6$ & $11.7\pm6.5$ & $0.0001\pm0.0001$ & $49.47\pm2.46$ & 36.5000 \\
3 & 3 & 3 & $1.5821\pm0.0000$ & 0 & -           & -           & $0.0000\pm0.0001$ & $951.06\pm53.71$ & 53.0000 \\
\hline
\end{tabular}
}
\end{table}

\begin{table}[H]
\centering
\caption{Performance summary of QUDO and qudit QAOA for the single depot VRP. Reported values are mean $\pm$ standard deviation over 10 random starts.}
\label{tab:qudo_dqaoa_svrp_summary}
\resizebox{\columnwidth}{!}{
\large
\begin{tabular}{c c c c c c c c c c}
\hline
$p$ & $N$ & $V$
& AR 
& Reach (\%) 
& Steps to target 
& Evals to target 
& $P_{\text{valid}}$ 
& Time (s) 
& Best cost \\
\hline
1 & 3 & 2 & $1.0000\pm0.0000$ & 100 & $1.2\pm0.4$ & $1.2\pm0.4$ & $0.0580\pm0.0116$ & $0.21\pm0.08$ & 33.5000 \\
1 & 3 & 3 & $1.0000\pm0.0000$ & 100 & $1.6\pm0.8$ & $1.6\pm0.8$ & $0.0279\pm0.0055$ & $0.41\pm0.06$ & 33.5000 \\
1 & 4 & 2 & $1.0122\pm0.0365$ & 90  & $3.2\pm1.8$ & $3.2\pm1.8$ & $0.0212\pm0.0059$ & $1.67\pm0.20$ & 37.0000 \\
1 & 4 & 3 & $1.0176\pm0.0527$ & 90  & $6.4\pm6.5$ & $6.4\pm6.5$ & $0.0122\pm0.0055$ & $10.31\pm0.67$ & 37.0000 \\
1 & 5 & 2 & $1.0472\pm0.0590$ & 50  & $9.6\pm7.1$ & $9.6\pm7.1$ & $0.0080\pm0.0019$ & $42.95\pm2.65$ & 44.5000 \\
1 & 5 & 3 & $1.3169\pm0.1394$ & 0   & $12.0\pm0.0$ & $12.0\pm0.0$ & $0.0035\pm0.0007$ & $337.25\pm17.59$ & 51.5000 \\
\hline
2 & 3 & 2 & $1.0000\pm0.0000$ & 100 & $1.0\pm0.0$ & $1.0\pm0.0$ & $0.0519\pm0.0161$ & $0.50\pm0.05$ & 33.5000 \\
2 & 3 & 3 & $1.0000\pm0.0000$ & 100 & $1.8\pm1.0$ & $1.8\pm1.0$ & $0.0190\pm0.0060$ & $1.52\pm0.23$ & 33.5000 \\
2 & 4 & 2 & $1.0000\pm0.0000$ & 100 & $2.1\pm1.5$ & $2.1\pm1.5$ & $0.0222\pm0.0039$ & $5.18\pm0.44$ & 37.0000 \\
2 & 4 & 3 & $1.0122\pm0.0365$ & 90  & $6.9\pm6.5$ & $6.9\pm6.5$ & $0.0103\pm0.0016$ & $32.54\pm1.13$ & 37.0000 \\
2 & 5 & 2 & $1.0326\pm0.0652$ & 80  & $16.8\pm12.0$ & $16.8\pm12.0$ & $0.0086\pm0.0021$ & $136.69\pm3.42$ & 44.5000 \\
2 & 5 & 3 & $1.1584\pm0.0865$ & 10  & $23.0\pm0.0$ & $23.0\pm0.0$ & $0.0036\pm0.0010$ & $1057.40\pm21.10$ & 44.5000 \\
\hline
3 & 3 & 2 & $1.0000\pm0.0000$ & 100 & $1.2\pm0.4$ & $1.2\pm0.4$ & $0.0415\pm0.0169$ & $0.93\pm0.07$ & 33.5000 \\
3 & 3 & 3 & $1.0000\pm0.0000$ & 100 & $1.6\pm0.9$ & $1.6\pm0.9$ & $0.0244\pm0.0060$ & $2.48\pm0.15$ & 33.5000 \\
3 & 4 & 2 & $1.0000\pm0.0000$ & 100 & $3.2\pm1.8$ & $3.2\pm1.8$ & $0.0192\pm0.0027$ & $8.30\pm0.04$ & 37.0000 \\
3 & 4 & 3 & $1.0000\pm0.0000$ & 100 & $7.9\pm7.5$ & $7.9\pm7.5$ & $0.0103\pm0.0014$ & $50.98\pm0.04$ & 37.0000 \\
3 & 5 & 2 & $1.0225\pm0.0275$ & 60  & $14.8\pm11.4$ & $14.8\pm11.4$ & $0.0083\pm0.0013$ & $210.99\pm0.05$ & 44.5000 \\
3 & 5 & 3 & $1.1213\pm0.0844$ & 20  & $26.5\pm7.5$ & $26.5\pm7.5$ & $0.0037\pm0.0007$ & $1619.31\pm1.06$ & 44.5000 \\
\hline
\end{tabular}
}
\end{table}

\subsection{Multi Depot VRP}
The multi depot vehicle routing problem (MDVRP) generalizes the single-depot VRP by allowing vehicles to originate from and return to multiple depots, each serving a subset of customers \citep{Cordeau2001MDVRPReview}. From an optimization perspective, the introduction of multiple depots significantly increases the combinatorial complexity of the problem, as feasible solutions must simultaneously determine customer to vehicle assignments, depot to vehicle associations, and route ordering while preventing vehicle collisions and infeasible depot transitions. These additional structural constraints increase the resource requirements of binary encodings, further amplifying the scalability limitations observed in QUBO based formulations. Hence, the MDVRP provides a natural benchmark for assessing the advantages of d-ary encodings and qudit based quantum optimization methods. Table \ref{tab:mdvrp_formulation} summarizes the Hamiltonian formulations of MDVRP under both QUBO and QUDO. A detailed derivation of each Hamiltonian is provided in Appendix C. 

\begin{table}[H]
\centering
\caption{QUBO and QUDO Hamiltonian formulations of MDVRP.
Here $N$ is the number of customers, $D$ the number of depots, $V$ the total number of vehicles}
\label{tab:mdvrp_formulation}
\resizebox{\columnwidth}{!}{
\begin{tabular}{p{0.47\columnwidth} | p{0.47\columnwidth}}
\hline
\textbf{QUBO formulation} & \textbf{QUDO formulation} \\
\hline

\textbf{Decision variables} &
\textbf{Decision variables} \\[4pt]

Binary variables
\[
X_{i,j}\in\{0,1\}
\]
indicate whether node $i$ (customer or depot) occupies position $j$ in a cyclic
sequence of length $M$. &
Discrete variables
\[
v_j \in \{0,\dots,N+D-1\}
\]
directly encode the node occupying position $j$ in the routing sequence. \\[8pt]

\textbf{Hamiltonian} &
\textbf{Hamiltonian} \\[4pt]

\[
\begin{aligned}
H_{\mathrm{MDVRP}}^{\mathrm{QUBO}}
&=
\sum_{j,i,k} D_{i,k}\,
X_{i,j}X_{k,j+1} \\
&\quad
+ H_{\mathrm{pos}}
+ H_{\mathrm{cust}}
+ H_{\mathrm{dep}}\\
+ H_{\mathrm{adj}}
\end{aligned}
\]
The first term encodes routing cost, while the remaining terms enforce
position occupancy, customer uniqueness, depot multiplicity, and
depot adjacency constraints. &
\[
\begin{aligned}
H_{\mathrm{MDVRP}}^{\mathrm{QUDO}}
&=
\sum_{j} D_{v_j,v_{j+1}} \\
&\quad
+ H_{\mathrm{cust}}
+ H_{\mathrm{dep}}
+ H_{\mathrm{adj}}
\end{aligned}
\]
Routing cost is encoded directly via discrete variables, with feasibility
enforced using collision-based penalties. \\[8pt]

\textbf{Constraint handling} &
\textbf{Constraint handling} \\[4pt]

Feasibility enforced through dense one-hot and adjacency penalties,
leading to $\mathcal{O}((N{+}D)(N{+}V))$ binary variables and dense
quadratic couplings. &
Feasibility embedded at the variable level; only collision penalties
are required, using $M=N+V$ qudits of local dimension $N{+}D$. \\

\hline
\end{tabular}
}
\end{table}

\subsubsection{Resource Requirements}

The resource requirements of the multi-depot VRP differ between the QUBO and QUDO formulations due to the underlying encoding strategy. In the QUBO formulation, a one-hot encoding is used to assign nodes to sequence positions. For $N$ customers, $D$ depots, and $V=\sum_d V_d$ vehicles, this requires binary variables $X_{i,j}$ with $i\in\{0,\dots,N+D-1\}$ and $j\in\{1,\dots,N+V\}$, resulting in $(N+D)(N+V)$ qubits. The corresponding Hilbert space grows exponentially as $2^{(N+D)(N+V)}$, quickly rendering both simulation and hardware implementation impractical.

The QUDO formulation replaces each block of one-hot variables with a single discrete variable $v_j$ that directly stores the node label. This requires only $N+V$ qudits, each of local dimension $d=N+D$. The total Hilbert space dimension is therefore $(N+D)^{(N+V)}$, representing an exponential compression in the number of quantum registers at the cost of a higher local dimension. This trade-off is particularly favorable for near-term quantum devices where connectivity and qubit count are limiting factors.

Table~\ref{tab:mdvrp_resources} summarizes the quantum resource requirements for representative multi-depot VRP instances. The number of qubits required by the QUBO formulation grows quadratically in problem size, while the QUDO formulation scales linearly in the number of sequence positions. This substantial reduction in quantum resources underpins the improved scalability observed in the qudit QAOA simulations.

% \begin{table}[t]
% \centering
% \caption{Quantum resource requirements for QUBO and QUDO formulations of the multi-depot VRP.}
% \label{tab:mdvrp_resources}
% \resizebox{\columnwidth}{!}{
% \begin{tabular}{cccc|cc|c}
% \hline
% $N$ & $D$ & $V$ 
% & QUBO qubits 
% & QUBO Hilbert space 
% & QUDO qudits 
% & QUDO Hilbert space \\ 
% \hline
% 3 & 2 & 2 
% & $(3+2)(3+2)=25$ 
% & $2^{25}\!\approx\!3.36\times10^{7}$ 
% & $5$ 
% & $5^{5}=3{,}125$ \\

% 4 & 2 & 2 
% & $(4+2)(4+2)=36$ 
% & $2^{36}\!\approx\!6.87\times10^{10}$ 
% & $6$ 
% & $6^{6}=46{,}656$ \\

% 4 & 3 & 3 
% & $(4+3)(4+3)=49$ 
% & $2^{49}\!\approx\!5.63\times10^{14}$ 
% & $7$ 
% & $7^{7}=823{,}543$ \\

% 5 & 3 & 3 
% & $(5+3)(5+3)=64$ 
% & $2^{64}\!\approx\!1.84\times10^{19}$ 
% & $8$ 
% & $8^{8}=16{,}777{,}216$ \\
% \hline
% \end{tabular}
% }
% \end{table}

\begin{table}[t]
\centering
\caption{Quantum resource requirements for QUBO and QUDO formulations of the multi-depot VRP.
Here $N$ is the number of customers, $D$ the number of depots, and $V$ the total number of vehicles.}
\label{tab:mdvrp_resources}
\resizebox{\columnwidth}{!}{
\begin{tabular}{ccc|c|c|c|c}
\hline
$N$ & $D$ & $V$
& \begin{tabular}[c]{@{}c@{}}QUBO qubits\\[-2pt]$((N{+}D)(N{+}V))$\end{tabular}
& \begin{tabular}[c]{@{}c@{}}QUBO Hilbert space\\[-2pt]$(2^{(N+D)(N+V)})$\end{tabular}
& \begin{tabular}[c]{@{}c@{}}QUDO qudits\\[-2pt]$(N{+}V)$\end{tabular}
& \begin{tabular}[c]{@{}c@{}}QUDO Hilbert space\\[-2pt]$((N{+}D)^{N+V})$\end{tabular} \\
\hline
3 & 2 & 2 & 25 & $2^{25}\!\approx\!3.36\times10^{7}$  & 5 & $5^{5}=3{,}125$ \\
4 & 2 & 2 & 36 & $2^{36}\!\approx\!6.87\times10^{10}$ & 6 & $6^{6}=46{,}656$ \\
4 & 3 & 3 & 49 & $2^{49}\!\approx\!5.63\times10^{14}$ & 7 & $7^{7}=823{,}543$ \\
5 & 3 & 3 & 64 & $2^{64}\!\approx\!1.84\times10^{19}$ & 8 & $8^{8}=16{,}777{,}216$ \\
\hline
\end{tabular}
}
\end{table}

% \begin{table}[t]
% \centering
% \caption{Quantum resource requirements for QUBO and QUDO formulations of the multi-depot VRP.
% Here $N$ is the number of customers, $D$ the number of depots, and $V$ the total number of vehicles.}
% \label{tab:mdvrp_resources}
% \resizebox{\columnwidth}{!}{
% \begin{tabular}{cccc|cc|cc}
% \hline
% $N$ & $D$ & $V$ 
% & QUBO qubits 
% & QUBO Hilbert space 
% & Formula 
% & QUDO qudits 
% & QUDO Hilbert space (Formula) \\ 
% \hline
% 3 & 2 & 2 
% & $(N{+}D)(N{+}V)=25$ 
% & $2^{25}\approx3.36\times10^{7}$ 
% & $2^{(N+D)(N+V)}$ 
% & $N{+}V=5$ 
% & $(N{+}D)^{N+V}=5^{5}=3{,}125$ \\

% 4 & 2 & 2 
% & $(N{+}D)(N{+}V)=36$ 
% & $2^{36}\approx6.87\times10^{10}$ 
% & $2^{(N+D)(N+V)}$ 
% & $N{+}V=6$ 
% & $(N{+}D)^{N+V}=6^{6}=46{,}656$ \\

% 4 & 3 & 3 
% & $(N{+}D)(N{+}V)=49$ 
% & $2^{49}\approx5.63\times10^{14}$ 
% & $2^{(N+D)(N+V)}$ 
% & $N{+}V=7$ 
% & $(N{+}D)^{N+V}=7^{7}=823{,}543$ \\

% 5 & 3 & 3 
% & $(N{+}D)(N{+}V)=64$ 
% & $2^{64}\approx1.84\times10^{19}$ 
% & $2^{(N+D)(N+V)}$ 
% & $N{+}V=8$ 
% & $(N{+}D)^{N+V}=8^{8}=16{,}777{,}216$ \\
% \hline
% \end{tabular}
% }
% \end{table}

\subsubsection{Experiment with QAOA and qudit QAOA}

\begin{table}[H]
\centering
\caption{Performance summary of QUBO and QAOA for the multi-depot VRP.
Reported values are mean $\pm$ standard deviation over 10 random starts.}
\label{tab:qubo_qaoa_mdvrp_summary}
\resizebox{\columnwidth}{!}{
\large
\begin{tabular}{c c c c c c c c c c c}
\hline
$p$ & $N$ & $D$ & $V$
& AR 
& Reach (\%) 
& Steps to target 
& Evals to target 
& $P_{\text{valid}}$ 
& Time (s) 
& Best cost \\
\hline
1 & 3 & 2 & 2 & -- & 0 & -- & -- & $0.0000\pm0.0000$ & $533.64\pm42.91$ & -- \\
\hline
2 & 3 & 2 & 2 & -- & 0 & -- & -- & $0.0000\pm0.0000$ & $1104.55\pm24.63$ & -- \\
\hline
3 & 3 & 2 & 2 & $1.0230\pm0.0000$ & 0 & -- & -- & $0.0000\pm0.0001$ & $1556.63\pm103.71$ & 44.5000 \\
\hline
\end{tabular}
}
\end{table}

\begin{table}[H]
\centering
\caption{Performance summary of QUDO and qudit QAOA for the multi-depot VRP.
Reported values are mean $\pm$ standard deviation over 10 random starts}
\label{tab:qudo_qaoa_mdvrp_summary}
\resizebox{\columnwidth}{!}{
\large
\begin{tabular}{c c c c c c c c c c c c}
\hline
$p$ & $N$ & $D$ & $V_d$ & $V$
& AR
& Reach (\%)
& Steps to target
& Evals to target
& $P_{\mathrm{valid}}$
& Time (s)
& Best cost \\
\hline

1 & 3 & 2 & {[}1,1{]} & 2
& $1.0000\pm0.0000$ & 100
& $1.0\pm0.0$ & $1.0\pm0.0$
& $0.0127\pm0.0000$ & $0.25\pm0.00$ & 43.5000 \\

1 & 3 & 3 & {[}1,1,1{]} & 3
& $1.0000\pm0.0000$ & 100
& $8.0\pm0.0$ & $8.0\pm0.0$
& $0.0010\pm0.0000$ & $12.95\pm0.00$ & 60.0000 \\

1 & 4 & 2 & {[}1,1{]} & 2
& $1.0638\pm0.0000$ & 0
& -- & --
& $0.0092\pm0.0000$ & $13.97\pm0.00$ & 50.0000 \\

1 & 4 & 3 & {[}1,1,1{]} & 3
& $1.2077\pm0.0000$ & 0
& $10.0\pm0.0$ & $10.0\pm0.0$
& $0.0005\pm0.0000$ & $299.72\pm0.00$ & 78.5000 \\

\hline

2 & 3 & 2 & {[}1,1{]} & 2
& $1.0000\pm0.0000$ & 100
& $1.0\pm0.0$ & $1.0\pm0.0$
& $0.0160\pm0.0000$ & $0.77\pm0.00$ & 43.5000 \\

2 & 3 & 3 & {[}1,1,1{]} & 3
& $1.0000\pm0.0000$ & 100
& $10.0\pm0.0$ & $10.0\pm0.0$
& $0.0013\pm0.0000$ & $41.06\pm0.00$ & 60.0000 \\

2 & 4 & 2 & {[}1,1{]} & 2
& $1.0000\pm0.0000$ & 100
& $1.0\pm0.0$ & $1.0\pm0.0$
& $0.0087\pm0.0000$ & $38.49\pm0.00$ & 47.0000 \\

2 & 4 & 3 & {[}1,1,1{]} & 3
& $1.0385\pm0.0000$ & 0
& -- & --
& $0.0010\pm0.0000$ & $967.60\pm0.00$ & 67.5000 \\

\hline

3 & 3 & 2 & {[}1,1{]} & 2
& $1.0000\pm0.0000$ & 100
& $1.0\pm0.0$ & $1.0\pm0.0$
& $0.0190\pm0.0000$ & $1.76\pm0.00$ & 43.5000 \\

3 & 3 & 3 & {[}1,1,1{]} & 3
& $1.0000\pm0.0000$ & 100
& $8.0\pm0.0$ & $8.0\pm0.0$
& $0.0048\pm0.0000$ & $60.22\pm0.00$ & 60.0000 \\

3 & 4 & 2 & {[}1,1{]} & 2
& $1.0000\pm0.0000$ & 100
& $5.0\pm0.0$ & $5.0\pm0.0$
& $0.0085\pm0.0000$ & $59.52\pm0.00$ & 47.0000 \\

3 & 4 & 3 & {[}1,1,1{]} & 3
& $1.0615\pm0.0000$ & 0
& -- & --
& $0.0010\pm0.0000$ & $1462.63\pm0.00$ & 69.0000 \\

\hline
\end{tabular}
}
\end{table}

Tables \ref{tab:qubo_qaoa_mdvrp_summary} and \ref{tab:qudo_qaoa_mdvrp_summary} summarize the performance of QAOA applied to the QUBO formulation and qudit QAOA applied to the QUDO formulation of MDVRP for circuit depths $p=1,2,3$. All reported metrics are averaged over 10 random initializations of the variational parameters. For small instances ($N=3$), the QUDO formulation consistently achieves an approximation ratio of unity and 100\% reach probability at depth $p=1$, indicating reliable recovery of the exact optimal solution with minimal circuit depth. As the number of depots increases, the steps-to-target and evaluations-to-target grow modestly, reflecting the increased combinatorial complexity introduced by additional depot constraints.

As the problem size increases to $N=4$, a clear separation between solvable and challenging instances emerges. For configurations with two depots, increasing the circuit depth improves robustness and enables convergence to the optimal solution with high reach probability. In contrast, instances with three depots exhibit approximation ratios greater than unity and zero reach probability across all tested depths, indicating that the optimizer fails to reach the exact optimum within the fixed computational budget. The approximation ratio remains well defined and close to unity, demonstrating that qudit QAOA still identifies high quality solutions even when exact convergence is not achieved. The observed increase in wall clock time with both $N$ and $D$ shows the exponential growth of the underlying qudit Hilbert space and the corresponding cost of classical parameter optimization.

Results for the corresponding QUBO formulation are intentionally limited for the MDVRP. The smallest instances exhibit vanishing feasibility probabilities and prohibitively large optimization times, with no reliable convergence to feasible solutions. These results highlight the practical advantage of QUDO encodings for structurally complex routing problems and motivate their application to other combinatorial optimization tasks, such as the unweighted Max-$K$-Cut problem considered next.

\section{Max-{K}-Cut (Unweighted)}

The unweighted Max-$K$-Cut problem is defined on an undirected graph $G=(V,E)$ with $|V|=N$ vertices. The objective is to partition the vertices into $K\ge 2$ disjoint subsets such that the number of edges whose endpoints belong to different subsets is maximized \citep{FriezeJerrum1997MaxKCut}. Table \ref{tab:maxkcut_formulation} summarizes the QUBO and QUDO formulation for the Max K cut problem. A detailed derivation of the Hamiltonian is provided in Appendix D.

\begin{table}[H]
\centering
\caption{QUBO and QUDO Hamiltonian formulations of the unweighted Max-$K$-Cut problem.}
\label{tab:maxkcut_formulation}
\renewcommand{\arraystretch}{1.15}
\resizebox{\columnwidth}{!}{
\begin{tabular}{p{0.48\columnwidth} | p{0.48\columnwidth}}
\hline
\textbf{QUBO formulation} & \textbf{QUDO formulation} \\
\hline

\textbf{Decision variables} & \textbf{Decision variables} \\[2pt]

One-hot binary variables:
\[
\begin{aligned}
x_{i,k}\in\{0,1\},\quad i=1,\dots,N,\; \\ k=1,\dots,K,
\end{aligned}
\]
$x_{i,k}=1$ if vertex $i$ is assigned to partition $k$. &
Single $K$-ary variable per vertex:
\[
s_i\in\{1,\dots,K\},\quad i=1,\dots,N,
\]
$s_i$ is the partition label of vertex $i$. \\[6pt]

\textbf{Hamiltonian} & \textbf{Hamiltonian} \\[2pt]

Cut objective:
\[
-\sum_{(i,j)\in E}\Big(1-\sum_{k=1}^{K}x_{i,k}x_{j,k}\Big)
\]
One-hot constraint penalty:
\[
A\sum_{i=1}^{N}\Big(\sum_{k=1}^{K}x_{i,k}-1\Big)^2
\]
Final QUBO:
\[
H^{\mathrm{QUBO}} = H_{\mathrm{cut}} + H_{\mathrm{onehot}}.
\]
&
Final QUDO:
\[
H^{\mathrm{QUDO}}
=
-\sum_{(i,j)\in E}\big(1-\delta(s_i=s_j)\big).
\]
(Equivalently, maximize edges with $s_i\neq s_j$.) \\

\hline
\end{tabular}
}
\end{table}

\subsection{Resource requirements}

The quantum resource requirements for Max-$K$-Cut differ substantially between the QUBO and QUDO formulations. In the QUBO encoding, each vertex must be assigned to exactly one of $K$ partitions using a one-hot representation, requiring $NK$ binary variables and additional quadratic penalty terms to enforce the assignment constraints. This leads to a Hilbert space of dimension $2^{NK}$ and dense couplings that grow rapidly with both the graph size and the number of partitions. In contrast, the QUDO formulation assigns a single $K$-ary variable to each vertex, naturally encoding the partition label without auxiliary constraints. As a result, the QUDO model requires only $N$ qudits of local dimension $K$, yielding a Hilbert space of size $K^{N}$ and eliminating the need for constraint penalties. This represents an exponential reduction in the number of quantum variables and a substantially simpler Hamiltonian structure, making the QUDO formulation particularly well suited for qudit QAOA implementations.

\begin{table}[H]
\centering
\caption{Quantum resource requirements for the unweighted Max-$K$-Cut problem under QUBO and QUDO formulations.}
\label{tab:maxkcut_resources}
\resizebox{\columnwidth}{!}{
\begin{tabular}{cc|c|c|c|c}
\hline
$N$ & $K$ 
& \begin{tabular}[c]{@{}c@{}}QUBO qubits\\[-2pt]$(NK)$\end{tabular}
& \begin{tabular}[c]{@{}c@{}}QUBO Hilbert space\\[-2pt]$(2^{NK})$\end{tabular}
& \begin{tabular}[c]{@{}c@{}}QUDO qudits\\[-2pt]$(N)$\end{tabular}
& \begin{tabular}[c]{@{}c@{}}QUDO Hilbert space\\[-2pt]$(K^{N})$\end{tabular} \\
\hline
5 & 2 & 10 & $2^{10}=1{,}024$ & 5 & $2^{5}=32$ \\
5 & 3 & 15 & $2^{15}=32{,}768$ & 5 & $3^{5}=243$ \\
6 & 3 & 18 & $2^{18}=262{,}144$ & 6 & $3^{6}=729$ \\
8 & 4 & 32 & $2^{32}\approx4.29\times10^{9}$ & 8 & $4^{8}=65{,}536$ \\
\hline
\end{tabular}
}
\end{table}
% \begin{table}[H]
% \centering
% \caption{Quantum resource requirements for the unweighted Max-$K$-Cut problem under QUBO and QUDO formulations.}
% \label{tab:maxkcut_resources}
% \resizebox{\columnwidth}{!}{
% \begin{tabular}{cc|cc|cc}
% \hline
% $N$ & $K$ 
% & QUBO qubits $(NK)$
% & QUBO Hilbert space $(2^{NK})$
% & QUDO qudits $(N)$
% & QUDO Hilbert space $(K^{N})$ \\
% \hline
% 5 & 2 
% & $5\times2=10$ 
% & $2^{10}=1{,}024$ 
% & $5$ 
% & $2^{5}=32$ \\

% 5 & 3 
% & $5\times3=15$ 
% & $2^{15}=32{,}768$ 
% & $5$ 
% & $3^{5}=243$ \\

% 6 & 3 
% & $6\times3=18$ 
% & $2^{18}=262{,}144$ 
% & $6$ 
% & $3^{6}=729$ \\

% 8 & 4 
% & $8\times4=32$ 
% & $2^{32}\approx4.29\times10^{9}$ 
% & $8$ 
% & $4^{8}=65{,}536$ \\
% \hline
% \end{tabular}
% }
% \end{table}

\subsection{Experiment with QAOA and qudit QAOA }
Tables \ref{tab:maxkcut_qubo_results} and \ref{tab:maxkcut_qudo_results} report the performance of QAOA applied to the QUBO formulation and qudit QAOA applied to the QUDO formulation of the unweighted Max-$K$-Cut problem, respectively, for circuit depths $p=1,2,3$. For each instance, we report the approximation ratio relative to the exact optimum, the reach probability, convergence statistics, and total wall-clock runtime, with all values averaged over 10 random initializations.

The QUBO based QAOA results exhibit rapidly deteriorating performance as either the number of vertices $N$ or the number of partitions $K$ increases. While near-optimal or optimal solutions are occasionally recovered for small instances and $K=2$, performance degrades substantially for larger $K$. This degradation results in reduced approximation ratios, sharply declining reach probabilities, and extremely low probabilities of sampling valid one-hot configurations. In addition, the wall-clock runtime increases dramatically, reaching several hours for instances with $N\geq6$ and $K=4$.

The QUDO based qudit QAOA results demonstrate consistently strong performance across all tested instances. For all values of $N$, $K$, and circuit depth $p$, the algorithm reliably achieves approximation ratios equal to one, with 100\% reach probability and near-unity valid-state sampling probability. Notably, these optimal results are obtained already at depth $p=1$ and persist for higher depths, while runtimes remain below one second even for the largest graphs considered. The optimization landscape is significantly smoother, and the search avoids the feasibility bottlenecks that dominate the QUBO formulation. These findings motivate the extension of QUDO encodings to other graph-based combinatorial optimization problems, including graph coloring, which we examine next.

\begin{table}[H]
\centering
\caption{QUBO and QAOA results for unweighted Max-$K$-Cut (10 multi-start runs per instance)}
\label{tab:maxkcut_qubo_results}
\resizebox{\columnwidth}{!}{
\large
\begin{tabular}{ccc|ccccccc}
\hline
$N$ & $K$ & $p$ 
& AR
& Reach(\%)
& Steps
& Evals
& $P_{\mathrm{valid}}$
& time (s)
& best\_cut \\
\hline
5 & 2 & 1 & $0.9750\pm0.0791$ & 90  & $2.2\pm1.9$  & $2.2\pm1.9$  & $0.1509\pm0.1195$ & $0.61\pm0.13$ & 4 \\
5 & 2 & 2 & $1.0000\pm0.0000$ & 100 & $2.3\pm1.8$  & $2.3\pm1.8$  & $0.1295\pm0.0772$ & $0.64\pm0.05$ & 4 \\
5 & 2 & 3 & $1.0000\pm0.0000$ & 100 & $1.8\pm1.2$  & $1.8\pm1.2$  & $0.1472\pm0.0900$ & $0.65\pm0.04$ & 4 \\
\hline
5 & 3 & 1 & $0.9429\pm0.0738$ & 60  & $6.6\pm5.5$  & $6.6\pm5.5$  & $0.0272\pm0.0306$ & $1.77\pm0.07$ & 7 \\
5 & 3 & 2 & $0.9429\pm0.1380$ & 80  & $6.6\pm6.1$  & $6.6\pm6.1$  & $0.0251\pm0.0321$ & $2.08\pm0.17$ & 7 \\
5 & 3 & 3 & $1.0000\pm0.0000$ & 100 & $5.2\pm4.3$  & $5.2\pm4.3$  & $0.0176\pm0.0086$ & $2.31\pm0.13$ & 7 \\
\hline
5 & 4 & 1 & $0.9500\pm0.1054$ & 80  & $4.8\pm3.6$  & $4.8\pm3.6$  & $0.0046\pm0.0058$ & $9.14\pm0.68$ & 4 \\
5 & 4 & 2 & $0.9167\pm0.1250$ & 60  & $6.6\pm7.2$  & $6.6\pm7.2$  & $0.0030\pm0.0029$ & $15.64\pm0.40$ & 4 \\
5 & 4 & 3 & $0.9250\pm0.1687$ & 80  & $6.4\pm5.6$  & $6.4\pm5.6$  & $0.0077\pm0.0087$ & $20.89\pm0.93$ & 4 \\
\hline
6 & 2 & 1 & $0.9833\pm0.0527$ & 90  & $4.2\pm4.0$  & $4.2\pm4.0$  & $0.0322\pm0.0326$ & $0.98\pm0.11$ & 6 \\
6 & 2 & 2 & $1.0000\pm0.0000$ & 100 & $4.9\pm2.9$  & $4.9\pm2.9$  & $0.0668\pm0.1117$ & $1.07\pm0.12$ & 6 \\
6 & 2 & 3 & $1.0000\pm0.0000$ & 100 & $3.1\pm2.6$  & $3.1\pm2.6$  & $0.1145\pm0.1736$ & $1.14\pm0.07$ & 6 \\
\hline
6 & 3 & 1 & $0.9000\pm0.0964$ & 40  & $8.4\pm6.1$  & $8.4\pm6.1$  & $0.0107\pm0.0121$ & $3.29\pm0.33$ & 7 \\
6 & 3 & 2 & $0.9143\pm0.0738$ & 40  & $5.4\pm4.3$  & $5.4\pm4.3$  & $0.0116\pm0.0215$ & $5.11\pm0.16$ & 7 \\
6 & 3 & 3 & $0.9000\pm0.0690$ & 30  & $14.2\pm8.4$ & $14.2\pm8.4$ & $0.0116\pm0.0087$ & $6.57\pm0.26$ & 7 \\
\hline
6 & 4 & 1 & $0.8143\pm0.1069$ & 10  & $3.0\pm0.0$  & $3.0\pm0.0$  & $0.0004\pm0.0006$ & $185.37\pm14.92$ & 10 \\
6 & 4 & 2 & $0.8714\pm0.0756$ & 10  & $12.8\pm7.5$ & $12.8\pm7.5$ & $0.0013\pm0.0017$ & $339.29\pm12.85$ & 10 \\
6 & 4 & 3 & $0.8300\pm0.1337$ & 10  & $5.5\pm6.4$  & $5.5\pm6.4$  & $0.0016\pm0.0014$ & $459.77\pm46.35$ & 10 \\
\hline
7 & 2 & 1 & $0.9375\pm0.0884$ & 60  & $9.3\pm4.8$  & $9.3\pm4.8$  & $0.0699\pm0.0793$ & $1.33\pm0.11$ & 8 \\
7 & 2 & 2 & $0.9375\pm0.0659$ & 50  & $12.6\pm9.4$ & $12.6\pm9.4$ & $0.0450\pm0.0696$ & $1.62\pm0.07$ & 8 \\
7 & 2 & 3 & $0.9375\pm0.0659$ & 50  & $12.9\pm8.1$ & $12.9\pm8.1$ & $0.0511\pm0.0459$ & $1.80\pm0.11$ & 8 \\
\hline
7 & 3 & 1 & $0.8857\pm0.1756$ & 50  & $10.8\pm4.2$ & $10.8\pm4.2$ & $0.0070\pm0.0102$ & $17.72\pm0.84$ & 7 \\
7 & 3 & 2 & $0.8429\pm0.0452$ & 0   & $17.5\pm7.2$ & $17.5\pm7.2$ & $0.0062\pm0.0103$ & $32.48\pm0.71$ & 6 \\
7 & 3 & 3 & $0.8429\pm0.1421$ & 30  & $13.8\pm10.0$& $13.8\pm10.0$& $0.0033\pm0.0030$ & $44.39\pm1.44$ & 7 \\
\hline
7 & 4 & 1 & $0.7556\pm0.2981$ & 20  & $13.5\pm7.8$ & $13.5\pm7.8$ & $0.0005\pm0.0007$ & $3221.57\pm320.05$ & 9 \\
7 & 4 & 2 & $0.7778\pm0.1217$ & 0 & $15.0\pm3.6$ & $15.0\pm3.6$ & $0.0008\pm0.0009$ & $6977.66\pm190.81$ & 8 \\
% 7 & 4 & 3 & -- & -- & -- & -- & -- & -- & -- \\
\hline
\end{tabular}
}
\end{table}

\begin{table}[H]
\centering
\caption{QUDO and qudit QAOA results for unweighted Max-$K$-Cut (10 multi start runs per instance)}
\label{tab:maxkcut_qudo_results}
\resizebox{\columnwidth}{!}{
\large
\begin{tabular}{ccc|ccccccc}
\hline
$N$ & $K$ & $p$ 
& AR 
& Reach(\%)
& Steps
& Evals
& $P_{\mathrm{valid}}$ 
& time (s)
& best\_cut \\
\hline
5 & 2 & 1 & $1.0000\pm0.0000$ & 100 & $1.0\pm0.0$ & $1.0\pm0.0$ & $1.0000\pm0.0000$ & $0.19\pm0.03$ & 4 \\
5 & 2 & 2 & $1.0000\pm0.0000$ & 100 & $1.0\pm0.0$ & $1.0\pm0.0$ & $1.0000\pm0.0000$ & $0.24\pm0.03$ & 4 \\
5 & 2 & 3 & $1.0000\pm0.0000$ & 100 & $1.0\pm0.0$ & $1.0\pm0.0$ & $1.0000\pm0.0000$ & $0.27\pm0.03$ & 4 \\
\hline
5 & 3 & 1 & $1.0000\pm0.0000$ & 100 & $1.3\pm0.9$ & $1.3\pm0.9$ & $1.0000\pm0.0000$ & $0.19\pm0.01$ & 7 \\
5 & 3 & 2 & $1.0000\pm0.0000$ & 100 & $1.0\pm0.0$ & $1.0\pm0.0$ & $1.0000\pm0.0000$ & $0.24\pm0.03$ & 7 \\
5 & 3 & 3 & $1.0000\pm0.0000$ & 100 & $1.2\pm0.6$ & $1.2\pm0.6$ & $1.0000\pm0.0000$ & $0.28\pm0.03$ & 7 \\
\hline
5 & 4 & 1 & $1.0000\pm0.0000$ & 100 & $1.0\pm0.0$ & $1.0\pm0.0$ & $1.0000\pm0.0000$ & $0.18\pm0.02$ & 4 \\
5 & 4 & 2 & $1.0000\pm0.0000$ & 100 & $1.0\pm0.0$ & $1.0\pm0.0$ & $1.0000\pm0.0000$ & $0.23\pm0.04$ & 4 \\
5 & 4 & 3 & $1.0000\pm0.0000$ & 100 & $1.0\pm0.0$ & $1.0\pm0.0$ & $1.0000\pm0.0000$ & $0.26\pm0.04$ & 4 \\
\hline
6 & 2 & 1 & $1.0000\pm0.0000$ & 100 & $1.2\pm0.6$ & $1.2\pm0.6$ & $1.0000\pm0.0000$ & $0.19\pm0.02$ & 6 \\
6 & 2 & 2 & $1.0000\pm0.0000$ & 100 & $1.0\pm0.0$ & $1.0\pm0.0$ & $1.0000\pm0.0000$ & $0.24\pm0.01$ & 6 \\
6 & 2 & 3 & $1.0000\pm0.0000$ & 100 & $1.0\pm0.0$ & $1.0\pm0.0$ & $1.0000\pm0.0000$ & $0.28\pm0.02$ & 6 \\
\hline
6 & 3 & 1 & $1.0000\pm0.0000$ & 100 & $1.2\pm0.6$ & $1.2\pm0.6$ & $1.0000\pm0.0000$ & $0.19\pm0.01$ & 7 \\
6 & 3 & 2 & $1.0000\pm0.0000$ & 100 & $1.0\pm0.0$ & $1.0\pm0.0$ & $1.0000\pm0.0000$ & $0.25\pm0.01$ & 7 \\
6 & 3 & 3 & $1.0000\pm0.0000$ & 100 & $1.0\pm0.0$ & $1.0\pm0.0$ & $1.0000\pm0.0000$ & $0.31\pm0.03$ & 7 \\
\hline
6 & 4 & 1 & $1.0000\pm0.0000$ & 100 & $1.0\pm0.0$ & $1.0\pm0.0$ & $1.0000\pm0.0000$ & $0.29\pm0.04$ & 10 \\
6 & 4 & 2 & $1.0000\pm0.0000$ & 100 & $1.0\pm0.0$ & $1.0\pm0.0$ & $1.0000\pm0.0000$ & $0.45\pm0.03$ & 10 \\
6 & 4 & 3 & $1.0000\pm0.0000$ & 100 & $1.1\pm0.3$ & $1.1\pm0.3$ & $1.0000\pm0.0000$ & $0.63\pm0.04$ & 10 \\
\hline
7 & 2 & 1 & $1.0000\pm0.0000$ & 100 & $1.0\pm0.0$ & $1.0\pm0.0$ & $1.0000\pm0.0000$ & $0.23\pm0.04$ & 8 \\
7 & 2 & 2 & $1.0000\pm0.0000$ & 100 & $1.1\pm0.3$ & $1.1\pm0.3$ & $1.0000\pm0.0000$ & $0.27\pm0.03$ & 8 \\
7 & 2 & 3 & $1.0000\pm0.0000$ & 100 & $1.2\pm0.4$ & $1.2\pm0.4$ & $1.0000\pm0.0000$ & $0.31\pm0.02$ & 8 \\
\hline
7 & 3 & 1 & $1.0000\pm0.0000$ & 100 & $1.0\pm0.0$ & $1.0\pm0.0$ & $1.0000\pm0.0000$ & $0.22\pm0.03$ & 7 \\
7 & 3 & 2 & $1.0000\pm0.0000$ & 100 & $1.2\pm0.6$ & $1.2\pm0.6$ & $1.0000\pm0.0000$ & $0.30\pm0.03$ & 7 \\
7 & 3 & 3 & $1.0000\pm0.0000$ & 100 & $1.0\pm0.0$ & $1.0\pm0.0$ & $1.0000\pm0.0000$ & $0.34\pm0.02$ & 7 \\
\hline
7 & 4 & 1 & $1.0000\pm0.0000$ & 100 & $1.0\pm0.0$ & $1.0\pm0.0$ & $1.0000\pm0.0000$ & $0.45\pm0.01$ & 9 \\
7 & 4 & 2 & $1.0000\pm0.0000$ & 100 & $1.0\pm0.0$ & $1.0\pm0.0$ & $1.0000\pm0.0000$ & $0.90\pm0.06$ & 9 \\
% 7 & 4 & 3 & -- & -- & -- & -- & -- & -- & -- \\
\hline
\end{tabular}
}
\end{table}

\section{Graph Coloring}
Given an undirected graph $G=(V,E)$ with $|V|=N$, the (proper) $K$-coloring problem asks
for an assignment of one of $K$ colors to each vertex such that adjacent vertices do not
share the same color. In the decision version, one asks whether a proper $K$-coloring exists \citep{graphcolor}.
In the optimization version considered here, constraint violations are penalized so that
ground states correspond to proper colorings when they exist. Table \ref{tab:graphcolor_formulation} summarizes the QUBO and QUDO Hamiltonian for the graph coloring problem. A detailed derivation of the Hamiltonian is provided in Appendix E.

\begin{table}[H]
\centering
\caption{QUBO and QUDO Hamiltonian formulations of the graph coloring problem.}
\label{tab:graphcolor_formulation}
\resizebox{\columnwidth}{!}{
\large
\begin{tabular}{p{0.48\columnwidth} | p{0.48\columnwidth}}
\hline
\textbf{QUBO formulation} & \textbf{QUDO formulation} \\
\hline

\textbf{Decision variables} & \textbf{Decision variables} \\[4pt]

Binary one-hot variables
\[
\begin{aligned}
x_{i,k} &\in \{0,1\} \\
i &\in \{1,\dots,N\} \\
k &\in \{1,\dots,K\}
\end{aligned}
\]
indicate whether vertex $i$ is assigned color $k$.
&
Discrete variables
\[
\begin{aligned}
s_i &\in \{1,\dots,K\} \\
i &\in \{1,\dots,N\}
\end{aligned}
\]
directly encode the color of vertex $i$. \\[10pt]

\textbf{Hamiltonian} & \textbf{Hamiltonian} \\[4pt]

\[
\begin{aligned}
H_{\mathrm{GC}}^{\mathrm{QUBO}} =\;&
A \sum_{i=1}^{N}\left(\sum_{k=1}^{K} x_{i,k}-1\right)^2 \\
&+ B \sum_{(i,j)\in E}\sum_{k=1}^{K} x_{i,k}\,x_{j,k}
\end{aligned}
\]
&
\[
\begin{aligned}
H_{\mathrm{GC}}^{\mathrm{QUDO}} =\;&
B \sum_{(i,j)\in E} \delta(s_i=s_j)
\end{aligned}
\]
\\[10pt]

\textbf{Constraint handling} & \textbf{Constraint handling} \\[4pt]

Vertex--color assignment is enforced using quadratic one-hot penalties. Invalid colorings can appear at low energy if penalty weights are insufficient.
&
Feasibility is native. Each variable always represents exactly one color. Only monochromatic edges are penalized. \\

\hline
\end{tabular}
}
\end{table}

% \begin{table}[H]
% \centering
% \caption{QUBO and QUDO Hamiltonian formulations of the graph coloring problem.}
% \label{tab:graphcolor_formulation}
% \resizebox{\columnwidth}{!}{
% \large
% \begin{tabular}{p{0.48\columnwidth} | p{0.48\columnwidth}}
% \hline
% \textbf{QUBO formulation} & \textbf{QUDO formulation} \\
% \hline

% \textbf{Decision variables} & \textbf{Decision variables} \\[2pt]

% Binary one-hot variables
% $
% x_{i,k}\in\{0,1\}
% $
% indicate whether vertex $i$ is assigned color $k$. &
% Discrete variables
% $
% s_i \in \{1,\dots,K\}
% $
% directly encode the color of vertex $i$. \\[8pt]

% \textbf{Hamiltonian} & \textbf{Hamiltonian} \\[2pt]

% $
% \displaystyle
% H_{\mathrm{GC}}^{\mathrm{QUBO}}
% =
% A \sum_{i=1}^{N}\left(\sum_{k=1}^{K}x_{i,k}-1\right)^2 + B \sum_{(i,j)\in E}\sum_{k=1}^{K} x_{i,k}x_{j,k}
% $
% &
% $
% \displaystyle
% H_{\mathrm{GC}}^{\mathrm{QUDO}}
% =
% B \sum_{(i,j)\in E} \delta(s_i=s_j)
% $
% \\[8pt]

% \textbf{Constraint handling} & \textbf{Constraint handling} \\[2pt]

% Vertex–color assignment enforced via quadratic one-hot penalties; invalid colorings may appear at low energy if penalties are insufficient. &
% Feasibility is native: each variable always represents exactly one color; only monochromatic edges are penalized. \\

% \hline
% \end{tabular}
% }
% \end{table}

\subsection{Resource Requirements}

Table~\ref{tab:graph_coloring_resources} shows the difference in quantum resource requirements between QUBO and QUDO encodings for graph coloring. In the QUBO formulation, each vertex–color assignment is represented using one-hot binary variables, leading to $NK$ qubits and a Hilbert space that grows as $2^{NK}$. Even for modest problem sizes, this results in an exponentially large state space, reaching over $10^{7}$ basis states for $N=8$ and $K=3$. In contrast, the QUDO formulation encodes the color of each vertex directly into a single $K$-level qudit, requiring only $N$ qudits and yielding a Hilbert space of dimension $K^{N}$. As shown in the table, this reduces the effective search space by several orders of magnitude for the same problem instance. This exponential compression of the Hilbert space provides a clear explanation for the improved feasibility and runtime performance observed for qudit QAOA relative to binary QAOA in graph coloring problems.

% \begin{table}[t]
% \centering
% \caption{Quantum resource requirements for QUBO and QUDO formulations of the graph coloring problem with $N$ vertices and $K$ colors.}
% \label{tab:graph_coloring_resources}
% \resizebox{\columnwidth}{!}{
% \begin{tabular}{cc|cc|cc}
% \hline
% $N$ & $K$ 
% & QUBO qubits 
% & QUBO Hilbert space 
% & QUDO qudits 
% & QUDO Hilbert space \\
% \hline
% 5 & 3 
% & $5\times3=15$ 
% & $2^{15}=32{,}768$ 
% & $5$ 
% & $3^{5}=243$ \\

% 6 & 3 
% & $6\times3=18$ 
% & $2^{18}=262{,}144$ 
% & $6$ 
% & $3^{6}=729$ \\

% 7 & 3 
% & $7\times3=21$ 
% & $2^{21}=2{,}097{,}152$ 
% & $7$ 
% & $3^{7}=2{,}187$ \\

% 8 & 3 
% & $8\times3=24$ 
% & $2^{24}=16{,}777{,}216$ 
% & $8$ 
% & $3^{8}=6{,}561$ \\
% \hline
% \end{tabular}
% }
% \end{table}

\begin{table}[H]
\centering
\caption{Quantum resource requirements for QUBO and QUDO formulations of the graph coloring problem with $N$ vertices and $K$ colors.}
\label{tab:graph_coloring_resources}
\resizebox{\columnwidth}{!}{
\begin{tabular}{cc|c|c|c|c}
\hline
$N$ & $K$ 
& \begin{tabular}[c]{@{}c@{}}QUBO qubits\\[-2pt]$(NK)$\end{tabular}
& \begin{tabular}[c]{@{}c@{}}QUBO Hilbert space\\[-2pt]$(2^{NK})$\end{tabular}
& \begin{tabular}[c]{@{}c@{}}QUDO qudits\\[-2pt]$(N)$\end{tabular}
& \begin{tabular}[c]{@{}c@{}}QUDO Hilbert space\\[-2pt]$(K^{N})$\end{tabular} \\
\hline
5 & 3 & 15 & $2^{15}=32{,}768$ & 5 & $3^{5}=243$ \\
6 & 3 & 18 & $2^{18}=262{,}144$ & 6 & $3^{6}=729$ \\
7 & 3 & 21 & $2^{21}=2{,}097{,}152$ & 7 & $3^{7}=2{,}187$ \\
8 & 3 & 24 & $2^{24}=16{,}777{,}216$ & 8 & $3^{8}=6{,}561$ \\
\hline
\end{tabular}
}
\end{table}

% \begin{table}[H]
% \centering
% \caption{Quantum resource requirements for QUBO and QUDO formulations of the graph coloring problem with $N$ vertices and $K$ colors. The table reports both the analytic scaling expressions and the corresponding numerical values.}
% \label{tab:graph_coloring_resources}
% \resizebox{\columnwidth}{!}{
% \begin{tabular}{cc|cc|cc}
% \hline
% $N$ & $K$ 
% & QUBO qubits ($NK$) 
% & QUBO Hilbert space ($2^{NK}$) 
% & QUDO qudits ($N$) 
% & QUDO Hilbert space ($K^{N}$) \\
% \hline
% 5 & 3 
% & $5\times3=15$ 
% & $2^{15}=32{,}768$ 
% & $5$ 
% & $3^{5}=243$ \\

% 6 & 3 
% & $6\times3=18$ 
% & $2^{18}=262{,}144$ 
% & $6$ 
% & $3^{6}=729$ \\

% 7 & 3 
% & $7\times3=21$ 
% & $2^{21}=2{,}097{,}152$ 
% & $7$ 
% & $3^{7}=2{,}187$ \\

% 8 & 3 
% & $8\times3=24$ 
% & $2^{24}=16{,}777{,}216$ 
% & $8$ 
% & $3^{8}=6{,}561$ \\
% \hline
% \end{tabular}
% }
% \end{table}

\subsection{Experiment with QAOA and qudit QAOA}

\begin{table}[H]
\centering
\caption{QUBO and QAOA performance for the Graph Coloring Problem. Reported values are mean $\pm$ standard deviation over 10 random starts.}
\label{tab:qubo_results}
\resizebox{\columnwidth}{!}{
\large
\begin{tabular}{cc|ccccccc}
\hline
$N$ & $p$
& AR
& Reach (\%)
& Steps
& Evals
& $P_{\mathrm{valid}}$
& time (s)
& best\_conf \\
\hline
5 & 1 & $1.0000\pm0.0000$ & 70 & $2.9\pm2.2$  & $2.9\pm2.2$  & $0.0222\pm0.0288$ & $0.79\pm0.05$ & 0 \\
6 & 1 & $1.0000\pm0.0000$ & 20 & $4.5\pm3.5$  & $4.5\pm3.5$  & $0.0180\pm0.0264$ & $2.53\pm0.23$ & 0 \\
7 & 1 & $1.0000\pm0.0000$ & 50 & $14.7\pm4.1$ & $14.7\pm4.1$ & $0.0125\pm0.0246$ & $16.22\pm0.97$ & 0 \\
8 & 1 & $1.0000\pm0.0000$ & 10 & $7.0\pm0.0$  & $7.0\pm0.0$  & $0.0068\pm0.0134$ & $144.39\pm8.88$ & 0 \\
\hline
5 & 2 & $1.0000\pm0.0000$ & 90 & $4.7\pm3.2$  & $4.7\pm3.2$  & $0.0407\pm0.0643$ & $1.06\pm0.04$ & 0 \\
6 & 2 & $1.0000\pm0.0000$ & 70 & $13.0\pm6.4$ & $13.0\pm6.4$ & $0.0221\pm0.0279$ & $4.35\pm0.19$ & 0 \\
7 & 2 & $1.0000\pm0.0000$ & 80 & $20.6\pm4.4$ & $20.6\pm4.4$ & $0.0099\pm0.0079$ & $28.70\pm0.82$ & 0 \\
8 & 2 & - & 0 & - & - & $0.0021\pm0.0018$ & $307.37\pm6.12$ & 1 \\
\hline
5 & 3 & $1.0000\pm0.0000$ & 80 & $3.8\pm3.2$  & $3.8\pm3.2$  & $0.0444\pm0.0685$ & $1.23\pm0.10$ & 0 \\
6 & 3 & $1.0000\pm0.0000$ & 50 & $6.5\pm4.5$  & $6.5\pm4.5$  & $0.0170\pm0.0179$ & $6.12\pm0.23$ & 0 \\
7 & 3 & $1.0000\pm0.0000$ & 40 & $13.5\pm8.4$ & $13.5\pm8.4$ & $0.0200\pm0.0398$ & $38.77\pm2.63$ & 0 \\
8 & 3 & $1.0000\pm0.0000$ & 10 & $24.0\pm0.0$ & $24.0\pm0.0$ & $0.0022\pm0.0019$ & $443.45\pm14.57$ & 0 \\
\hline
\end{tabular}
}
\end{table}

\begin{table}[H]
\centering
\caption{QUDO and qudit QAOA performance for the Graph Coloring problem. Reported values are mean $\pm$ standard deviation over 10 random starts.}
\label{tab:qudo_results}
\resizebox{\columnwidth}{!}{
\large
\begin{tabular}{cc|ccccccc}
\hline
$N$ & $p$
& AR
& Reach (\%)
& Steps
& Evals
& $P_{\mathrm{valid}}$
& time (s)
& best\_conf \\
\hline
5 & 1 & $1.0000\pm0.0000$ & 100 & $1.0\pm0.0$ & $1.0\pm0.0$ & $1.0000\pm0.0000$ & $0.19\pm0.01$ & 0 \\
6 & 1 & $1.0000\pm0.0000$ & 100 & $1.0\pm0.0$ & $1.0\pm0.0$ & $1.0000\pm0.0000$ & $0.23\pm0.03$ & 0 \\
7 & 1 & $1.0000\pm0.0000$ & 100 & $1.0\pm0.0$ & $1.0\pm0.0$ & $1.0000\pm0.0000$ & $0.27\pm0.03$ & 0 \\
8 & 1 & $1.0000\pm0.0000$ & 100 & $1.2\pm0.4$ & $1.2\pm0.4$ & $1.0000\pm0.0000$ & $0.40\pm0.01$ & 0 \\
\hline
5 & 2 & $1.0000\pm0.0000$ & 100 & $1.0\pm0.0$ & $1.0\pm0.0$ & $1.0000\pm0.0000$ & $0.23\pm0.02$ & 0 \\
6 & 2 & $1.0000\pm0.0000$ & 100 & $1.0\pm0.0$ & $1.0\pm0.0$ & $1.0000\pm0.0000$ & $0.30\pm0.00$ & 0 \\
7 & 2 & $1.0000\pm0.0000$ & 100 & $1.0\pm0.0$ & $1.0\pm0.0$ & $1.0000\pm0.0000$ & $0.32\pm0.03$ & 0 \\
8 & 2 & $1.0000\pm0.0000$ & 100 & $1.0\pm0.0$ & $1.0\pm0.0$ & $1.0000\pm0.0000$ & $0.65\pm0.06$ & 0 \\
\hline
5 & 3 & $1.0000\pm0.0000$ & 100 & $1.1\pm0.3$ & $1.1\pm0.3$ & $1.0000\pm0.0000$ & $0.26\pm0.03$ & 0 \\
6 & 3 & $1.0000\pm0.0000$ & 100 & $1.2\pm0.6$ & $1.2\pm0.6$ & $1.0000\pm0.0000$ & $0.34\pm0.03$ & 0 \\
7 & 3 & $1.0000\pm0.0000$ & 100 & $1.0\pm0.0$ & $1.0\pm0.0$ & $1.0000\pm0.0000$ & $0.36\pm0.04$ & 0 \\
8 & 3 & $1.0000\pm0.0000$ & 100 & $1.0\pm0.0$ & $1.0\pm0.0$ & $1.0000\pm0.0000$ & $0.93\pm0.03$ & 0 \\
\hline
\end{tabular}
}
\end{table}

Table~\ref{tab:qubo_results} summarizes the performance of QAOA applied to the QUBO formulation of the graph coloring problem. Although approximation ratios of unity are reported for most configurations where solutions are obtained, feasibility rapidly degrades with increasing problem size. The probability of sampling valid configurations decreases by over an order of magnitude as $N$ grows, and the reach percentage drops sharply, falling to 10\% for $N=8$ at $p=1$ and $p=3$. This loss of feasibility is accompanied by a substantial increase in optimization effort, with the mean number of steps and evaluations rising to double digits for larger instances. In some cases, such as $(N,p)=(8,2)$, no target-achieving solution is found within the allotted budget, leading to undefined steps and approximation ratios despite nontrivial runtime.

Table~\ref{tab:qudo_results} reports the corresponding results for qudit QAOA applied to the QUDO formulation. In contrast, the QUDO encoding achieves AR$=1.0000$ with 100\% reach across all tested graph sizes and circuit depths. Feasibility is guaranteed by construction, yielding $P_{\mathrm{valid}}=1.0000$ in every case and enabling convergence to the optimum in approximately one optimization step on average. As a result, wall-clock times remain below one second even for the largest instances considered.

Taken together, these results show that the primary limitation of the QUBO formulation for graph coloring is not objective quality but feasibility enforcement via one-hot penalties, which leads to vanishing valid-sample probabilities and poor scaling. By embedding feasibility directly into the variable domain, the QUDO formulation produces a smoother optimization landscape and stable performance across system sizes. This motivates the application of d-ary encodings to problems with more complex structural constraints, such as job scheduling, which we consider next.

\section{Job Scheduling Problem}
Job scheduling problems arise in manufacturing, computing, and logistics, where a set of jobs must be processed subject to limited resources. Even in simplified forms, it is NP-Hard in nature and serve as benchmarks for combinatorial optimization. In this work, we consider the single-machine job scheduling problem, where $N$ jobs must be processed sequentially on a single machine \citep{ConwayMaxwellMiller1967Scheduling}.

Each job $i \in \{1,\dots,N\}$ is characterized by a processing time $p_i > 0$ and a weight $w_i > 0$. The objective is to determine an ordering of jobs that minimizes the total weighted completion time. Table \ref{tab:jobsched_formulation} summarizes the Hamiltonian for both the QUBO and QUDO encodings. A detailed derivation of the Hamiltonian is provided in Appendix F.

\begin{table}[H]
\centering
\caption{QUBO and QUDO Hamiltonian formulations of the job scheduling problem.}
\label{tab:jobsched_formulation}
\resizebox{\columnwidth}{!}{
\large
\begin{tabular}{p{0.48\columnwidth} | p{0.48\columnwidth}}
\hline
\textbf{QUBO formulation} & \textbf{QUDO formulation} \\
\hline

\textbf{Decision variables} & \textbf{Decision variables} \\[4pt]

Binary one-hot variables
\[
\begin{aligned}
x_{i,j} &\in \{0,1\} \\
i &\in \{1,\dots,N\} \\
j &\in \{1,\dots,N\}
\end{aligned}
\]
indicate whether job $i$ is scheduled at position $j$.
&
Discrete variables
\[
\begin{aligned}
v_j &\in \{1,\dots,N\} \\
j &\in \{1,\dots,N\}
\end{aligned}
\]
denote the job scheduled at position $j$. \\[10pt]

\textbf{Hamiltonian} & \textbf{Hamiltonian} \\[4pt]

{\small
\[
\begin{aligned}
H_{\mathrm{sched}}^{\mathrm{QUBO}} =\;&
\sum_{j,i} w_i\, p_i\, j\, x_{i,j} \\
&+ A \sum_{j}\Big(\sum_{i} x_{i,j}-1\Big)^2 \\
&+ A \sum_{i}\Big(\sum_{j} x_{i,j}-1\Big)^2
\end{aligned}
\]
}
&
{\small
\[
\begin{aligned}
H_{\mathrm{sched}}^{\mathrm{QUDO}} =\;&
\sum_{j} w_{v_j}\, p_{v_j}\, j \\
&+ A \sum_{p<q}\sum_{i}
\delta(v_p=i)\,\delta(v_q=i)
\end{aligned}
\]
}
\\[10pt]

\textbf{Constraint handling} & \textbf{Constraint handling} \\[4pt]

Permutation constraints are enforced via quadratic one-hot penalties. This leads to dense couplings and many infeasible low-energy states.
&
Feasibility is enforced through collision penalties. Each position always contains exactly one job by construction. \\

\hline
\end{tabular}
}
\end{table}

% \begin{table}[H]
% \centering
% \caption{QUBO and QUDO Hamiltonian formulations of the job scheduling problem.}
% \label{tab:jobsched_formulation}
% \resizebox{\columnwidth}{!}{
% \large
% \begin{tabular}{p{0.48\columnwidth} | p{0.48\columnwidth}}
% \hline
% \textbf{QUBO formulation} & \textbf{QUDO formulation} \\
% \hline

% \textbf{Decision variables} & \textbf{Decision variables} \\[2pt]

% Binary one-hot variables
% $
% x_{i,j}\in\{0,1\}
% $
% indicate whether job $i$ is scheduled at position $j$. &
% Discrete variables
% $
% v_j \in \{1,\dots,N\}
% $
% denote the job scheduled at position $j$. \\[8pt]

% \textbf{Hamiltonian} & \textbf{Hamiltonian} \\[2pt]

% $
% \displaystyle
% H_{\mathrm{sched}}^{\mathrm{QUBO}}
% =
% \sum_{j,i} w_i p_i j\, x_{i,j}
% +
% A \sum_{j}\Big(\sum_{i} x_{i,j}-1\Big)^2
% +
% A \sum_{i}\Big(\sum_{j} x_{i,j}-1\Big)^2
% $
% &
% $
% \displaystyle
% H_{\mathrm{sched}}^{\mathrm{QUDO}}
% =
% \sum_{j} w_{v_j} p_{v_j} j
% +
% A \sum_{p<q}\sum_{i}
% \delta(v_p=i)\delta(v_q=i)
% $
% \\[8pt]

% \textbf{Constraint handling} & \textbf{Constraint handling} \\[2pt]

% Permutation constraints enforced via quadratic one-hot penalties, leading to dense couplings and many infeasible low-energy states. &
% Feasibility enforced through collision penalties; each position always contains exactly one job by construction. \\

% \hline
% \end{tabular}
% }
% \end{table}

\subsection{Resource Requirements}
Table~\ref{tab:job_scheduling_resources} compares the quantum resource requirements of the job scheduling problem under QUBO and QUDO formulations. In the standard QUBO encoding, one-hot binary variables are introduced to represent the assignment of each job to each time slot. This requires $N\times T$ qubits for $N$ jobs and $T$ discrete time slots, leading to an exponentially large Hilbert space of dimension $2^{NT}$. The QUDO formulation replaces each one-hot block with a single discrete variable, such that each job is represented by one qudit of local dimension $T$. This reduces the hardware requirement to $N$ qudits and yields a Hilbert space of size $T^{N}$. As shown in the table, this change results in an exponential compression of the accessible search space even for modest problem sizes. This reduction provides a clear explanation for the improved feasibility and runtime performance observed for qudit-based QAOA in comparison to binary QAOA for the job scheduling problem.

\begin{table}[H]
\centering
\caption{Quantum resource requirements for the job scheduling problem under QUBO and QUDO formulations.
Here $N$ denotes the number of jobs and $T$ the number of discrete time slots.}
\label{tab:job_scheduling_resources}
\resizebox{\columnwidth}{!}{
\begin{tabular}{cc|c|c|c|c}
\hline
$N$ & $T$
& \begin{tabular}[c]{@{}c@{}}QUBO qubits\\[-2pt]$(NT)$\end{tabular}
& \begin{tabular}[c]{@{}c@{}}QUBO Hilbert space\\[-2pt]$(2^{NT})$\end{tabular}
& \begin{tabular}[c]{@{}c@{}}QUDO qudits\\[-2pt]$(N)$\end{tabular}
& \begin{tabular}[c]{@{}c@{}}QUDO Hilbert space\\[-2pt]$(T^{N})$\end{tabular} \\
\hline
3 & 3 & 9  & $2^{9}=512$                         & 3 & $3^{3}=27$ \\
4 & 4 & 16 & $2^{16}=65{,}536$                   & 4 & $4^{4}=256$ \\
5 & 5 & 25 & $2^{25}\!\approx\!3.36\times10^{7}$ & 5 & $5^{5}=3{,}125$ \\
6 & 6 & 36 & $2^{36}\!\approx\!6.87\times10^{10}$& 6 & $6^{6}=46{,}656$ \\
\hline
\end{tabular}
}
\end{table}

\subsection{Experiment with QAOA and qudit QAOA}
\begin{table}[H]
\centering
\caption{QUBO, QAOA results for the job scheduling problem. Reported values are mean $\pm$ standard deviation over 10 random starts.}
\label{tab:jobsched_qubo_results}
\resizebox{\columnwidth}{!}{
\large
\begin{tabular}{cc|cccccc}
\hline
$N$ & $p$
& AR
& Reach (\%)
& Steps
& Evals
& $P_{\mathrm{valid}}$
& time (s) \\
\hline
3 & 1 & $1.0322\pm0.0915$ & 100 & $2.1\pm1.4$ & $2.1\pm1.4$ & $0.0160\pm0.0120$ & $0.92\pm0.07$ \\
4 & 1 & $1.7601\pm0.3738$ & 0   & --          & --          & $0.0003\pm0.0003$ & $4.95\pm0.27$ \\
5 & 1 & --                & 0   & --          & --          & $0.0000\pm0.0000$ & $999.24\pm75.61$ \\
\hline
3 & 2 & $1.0000\pm0.0000$ & 100 & $1.4\pm1.0$ & $1.3\pm0.7$ & $0.0153\pm0.0075$ & $1.07\pm0.03$ \\
4 & 2 & $1.5057\pm0.3932$ & 30  & $18.7\pm10.6$ & $18.0\pm10.5$ & $0.0004\pm0.0005$ & $6.41\pm0.34$ \\
5 & 2 & --                & 0   & --          & --          & $0.0000\pm0.0000$ & $1945.57\pm80.98$ \\
\hline
3 & 3 & $1.0015\pm0.0047$ & 100 & $3.1\pm4.5$ & $2.7\pm3.5$ & $0.0163\pm0.0100$ & $1.17\pm0.03$ \\
4 & 3 & $1.5920\pm0.4072$ & 20  & $14.0\pm12.7$ & $12.5\pm12.0$ & $0.0005\pm0.0004$ & $7.44\pm0.25$ \\
5 & 3 & --                & 0   & --          & --          & $0.0000\pm0.0000$ & $2769.44\pm166.94$ \\
\hline
\end{tabular}
}
\end{table}

\begin{table}[H]
\centering
\caption{QUDO, qudit QAOA results for the job scheduling problem. Reported values are mean $\pm$ standard deviation over 10 random starts.}
\label{tab:jobsched_qudo_results}
\resizebox{\columnwidth}{!}{
\large
\begin{tabular}{cc|cccccc}
\hline
$N$ & $p$
& AR
& Reach (\%)
& Steps
& Evals
& $P_{\mathrm{valid}}$
& time (s) \\
\hline
3 & 1 & $1.0000\pm0.0000$ & 100 & $1.1\pm0.3$ & $1.1\pm0.3$ & $0.2040\pm0.0522$ & $0.09\pm0.02$ \\
4 & 1 & $1.0000\pm0.0000$ & 100 & $1.6\pm1.3$ & $1.5\pm1.0$ & $0.0937\pm0.0224$ & $0.14\pm0.04$ \\
5 & 1 & $1.0000\pm0.0000$ & 100 & $8.5\pm5.2$ & $7.8\pm5.3$ & $0.0433\pm0.0081$ & $0.40\pm0.05$ \\
\hline
3 & 2 & $1.0000\pm0.0000$ & 100 & $1.1\pm0.3$ & $1.1\pm0.3$ & $0.1925\pm0.0514$ & $0.18\pm0.02$ \\
4 & 2 & $1.0000\pm0.0000$ & 100 & $1.8\pm1.0$ & $1.8\pm1.0$ & $0.1036\pm0.0282$ & $0.28\pm0.04$ \\
5 & 2 & $1.0000\pm0.0000$ & 100 & $3.6\pm2.9$ & $3.4\pm2.6$ & $0.0370\pm0.0057$ & $0.76\pm0.06$ \\
\hline
3 & 3 & $1.0000\pm0.0000$ & 100 & $1.1\pm0.3$ & $1.1\pm0.3$ & $0.2310\pm0.0793$ & $0.26\pm0.01$ \\
4 & 3 & $1.0000\pm0.0000$ & 100 & $1.2\pm0.4$ & $1.2\pm0.4$ & $0.0906\pm0.0205$ & $0.36\pm0.01$ \\
5 & 3 & $1.0000\pm0.0000$ & 100 & $10.5\pm8.4$ & $9.9\pm8.1$ & $0.0388\pm0.0054$ & $1.09\pm0.02$ \\
\hline
\end{tabular}
}
\end{table}

Tables~\ref{tab:jobsched_qubo_results} and~\ref{tab:jobsched_qudo_results} compare the performance of QAOA applied to the QUBO formulation and qudit QAOA applied to the QUDO formulation of the job scheduling problem across increasing numbers of jobs and circuit depths $p=1,2,3$.

For the QUBO formulation, performance deteriorates rapidly as the problem size increases. While near optimal schedules are consistently obtained for $N=3$, feasibility collapses for $N\geq4$, with the probability of sampling valid schedules dropping to $10^{-4}$ or below. For $N=5$, no feasible solution is found at any circuit depth within the allotted computational budget, resulting in undefined steps to target and approximation ratios. Increasing the circuit depth does not alleviate this issue; instead, it leads to substantial growth in wall-clock time without improving feasibility.

In contrast, the QUDO formulation exhibits stable and robust performance across all tested instances. Optimal schedules are recovered for all values of $N$ and $p$, with approximation ratios equal to unity and 100\% reach in every case. Feasibility probabilities remain orders of magnitude higher than in the QUBO formulation, and convergence typically occurs within only a few optimization steps. Although the number of evaluations increases modestly with problem size, total runtime remains below a few seconds even for $N=5$, demonstrating favorable scaling. 

% \section{Discussion}

% These results are obtained from noiseless simulations and are limited to modest system sizes due to classical simulation constraints. Nevertheless, the consistency of the observed trends across routing, graph, and scheduling problems indicates that the advantages of QUDO are general and stem from a fundamental reduction in constraint-driven complexity.

\section{Conclusion and Future Work}
In this work, we demonstrated that QUDO provides a powerful alternative to conventional QUBO formulations for combinatorial optimization problems. By encoding decision variables in higher dimensional Hilbert spaces, QUDO avoids the extensive penalty terms that often dominate QUBO formulations. Through systematic benchmarking on the Traveling Salesman Problem, two variants of the Vehicle Routing Problem, Max-K-Cut, graph coloring, and job scheduling, we showed that qudit based QAOA achieves superior scalability, higher feasibility, and more consistent recovery of optimal solutions at shallow circuit depths.

Our results indicate that the advantages of QUDO arise primarily from structural efficiency at the encoding level rather than increased circuit expressivity alone. Native constraint satisfaction and reduced Hilbert space growth lead to smoother optimization landscapes and significantly lower classical and quantum resource requirements. These findings highlight qudit-based variational algorithms as a promising pathway toward practical quantum advantage in constrained combinatorial optimization.

Several important directions follow from this work. A natural extension is the inclusion of additional practical constraints, such as vehicle capacity limits and time windows in routing problems, as well as precedence constraints in job scheduling. Studying such constraints will further test the expressive power of D-ary encodings and clarify their advantages over binary formulations.

While this study relies on classical simulation of qudit circuits, future work should also explore hardware oriented implementations of qudit QAOA. Platforms that naturally support higher-dimensional systems, such as trapped ions \citep{Shi2026QuditImplementation} and photonic qudits \citep{karacsony2023efficientquditbasedscheme}, are particularly promising. Investigating gate decompositions, noise resilience, and connectivity requirements for qudit mixers and cost unitaries will be critical for assessing near term feasibility.

Finally, the design of problem informed qudit mixer Hamiltonians remains largely unexplored. Tailoring mixers to exploit the structure of specific combinatorial problems may further improve convergence rates and solution quality beyond what is achievable with generic shift operators.

\newpage

% \printbibliography
%\subsection{Multi Depot, multi vehicle with constraints}

% \newpage

\section*{Appendix A}
\subsection*{Traveling Salesman Problem}
\subsubsection*{QUBO Formulation}

The standard QUBO formulation of the Traveling Salesman Problem introduces binary decision variables
\begin{equation}
X_{i,j} \in \{0,1\},
\qquad i,j \in \{1,\dots,N\},
\end{equation}
where $X_{i,j}=1$ indicates that city $i$ is visited at position $j$ in the tour. A valid TSP tour corresponds to a permutation matrix $X$, such that each city appears exactly once and each tour position is occupied by exactly one city.

\paragraph{Objective function}
The total tour length is expressed as
\begin{equation}
\min_{X}\;
\sum_{j=1}^{N}
\sum_{i=1}^{N}
\sum_{k=1}^{N}
D_{i,k}\,
X_{i,j}\,
X_{k,(j+1)\bmod N},
\end{equation}
which accumulates the distance between consecutive cities in the cyclic tour.

\paragraph{QUBO Hamiltonian}
The permutation constraints are enforced through quadratic penalty terms. The resulting QUBO energy function is given by
\begin{align}
E^{\mathrm{QUBO}}(X)
&=
\sum_{j=1}^{N}
\sum_{i=1}^{N}
\sum_{k=1}^{N}
D_{i,k}\,
X_{i,j}\,
X_{k,(j+1)\bmod N}
\nonumber \\
&\quad
+ A \sum_{j=1}^{N}
\left(
\sum_{i=1}^{N} X_{i,j} - 1
\right)^2
+ A \sum_{i=1}^{N}
\left(
\sum_{j=1}^{N} X_{i,j} - 1
\right)^2 .
\end{align}

The first penalty term enforces that exactly one city occupies each tour position, while the second ensures that each city appears exactly once in the tour. Although these constraints are quadratic, they induce dense couplings among the $N^2$ binary variables, resulting in a rapidly increasing Hamiltonian complexity as the problem size grows\citep{SmithMiles2025TSPQuantum, Lucas_2014}.

\subsubsection*{QUDO Formulation}

In contrast to the QUBO formulation, the Traveling Salesman Problem can be expressed using a native $D$-ary encoding, where each decision variable directly represents the city visited at a given tour position.

\paragraph{Decision variables}
We introduce $N$ discrete variables
\begin{equation}
v_j \in \{1,\dots,N\},
\qquad j \in \{1,\dots,N\},
\end{equation}
where $v_j$ denotes the city visited at position $j$ in the tour. Each variable $v_j$ is a $d$-level variable (qudit) with $d=N$ possible states.

\paragraph{Objective function}
The total tour length is expressed directly as
\begin{equation}
\min_{v}\;
\sum_{j=1}^{N}
D_{v_j,\,v_{(j+1)\bmod N}}
\end{equation}
which accumulates the distance between consecutive cities along the cyclic tour.

\paragraph{QUDO Hamiltonian}
To ensure that each city is visited exactly once, we introduce a quadratic penalty that suppresses repeated city assignments across different positions. Using indicator functions $\delta(v=a)$, defined as
\[
\delta(v=a)=
\begin{cases}
1, & \text{if } v=a,\\
0, & \text{otherwise},
\end{cases}
\]
the QUDO energy function is given by
\begin{align}
E^{\mathrm{QUDO}}(v)
&=
\sum_{j=1}^{N}
\sum_{a=1}^{N}
\sum_{b=1}^{N}
D_{a,b}\,
\delta(v_j=a)\,
\delta(v_{(j+1)\bmod N}=b)
\nonumber \\
&\quad
+ A
\sum_{1 \le p < q \le N}
\sum_{a=1}^{N}
\delta(v_p=a)\,
\delta(v_q=a).
\end{align}

The first term encodes the tour cost by coupling adjacent tour positions, while the second term enforces customer uniqueness by penalizing configurations in which the same city appears at multiple positions. Unlike the QUBO formulation, the QUDO encoding eliminates the need for position constraints entirely, as each tour position is represented by a single $d$-ary variable that always assumes exactly one valid city label. The native $d$-ary structure naturally aligns with qudit-based implementations of QAOA, where generalized shift and phase operators act directly on the multi-level decision variables.

\section*{Appendix B}
\subsection*{Single Depot VRP}
\subsubsection*{QUBO formulation}

We formulate the single-depot vehicle routing problem (VRP) with $N$ customers and $V$ identical vehicles as a quadratic unconstrained binary optimization (QUBO) problem. All vehicles start and end at a common depot labeled $0$. The routing solution is represented as a cyclic sequence of length $M = N + V$, where each occurrence of the depot acts as a route separator \citep{Borle2021VRP}.

\paragraph{Decision variables}
We introduce binary variables
\begin{equation}
X_{i,j} \in \{0,1\},
\quad
i \in \{0,1,\dots,N\},\;
j \in \{1,\dots,M\},
\end{equation}
where $X_{i,j} = 1$ if node $i$ (customer or depot) occupies position $j$ in the sequence.

\paragraph{Objective function}
The total routing cost is expressed as
\begin{equation}
H_{\mathrm{cost}}
=
\sum_{j=1}^{M}
\sum_{i=0}^{N}
\sum_{k=0}^{N}
D_{i,k}\,
X_{i,j}\,
X_{k,(j+1)\bmod M},
\end{equation}
which accounts for all customer--customer, depot--customer, and customer--depot transitions.

\paragraph{Constraints}
Feasibility is enforced through quadratic penalty terms. First, each position must be occupied by exactly one node:
\begin{equation}
H_{\mathrm{pos}}
=
A\sum_{j=1}^{M}
\left(
\sum_{i=0}^{N} X_{i,j} - 1
\right)^2.
\end{equation}

Second, each customer must be visited exactly once:
\begin{equation}
H_{\mathrm{city}}
=
A\sum_{i=1}^{N}
\left(
\sum_{j=1}^{M} X_{i,j} - 1
\right)^2.
\end{equation}

Third, the depot must appear exactly $V$ times in the sequence, corresponding to the number of vehicles:
\begin{equation}
H_{\mathrm{depot}}
=
B\left(
\sum_{j=1}^{M} X_{0,j} - V
\right)^2.
\end{equation}

\paragraph{QUBO Hamiltonian}
The complete QUBO Hamiltonian is therefore
\begin{equation}
H_{\mathrm{VRP}}^{\mathrm{QUBO}}
=
H_{\mathrm{cost}}
+
H_{\mathrm{pos}}
+
H_{\mathrm{city}}
+
H_{\mathrm{depot}},
\end{equation}
where the penalty weights satisfy $A \gg B > 0$ to ensure feasibility is prioritized over cost minimization.

Minimizing $H_{\mathrm{VRP}}^{\mathrm{QUBO}}$ yields a valid solution to the single-depot VRP. However, this formulation requires $(N+1)(N+V)$ binary variables and introduces dense quadratic couplings, which significantly limits scalability on near-term quantum hardware \citep{Lucas_2014}.

\subsubsection*{QUDO formulation}

We now formulate the single-depot VRP as a QUDO problem. Unlike the QUBO approach, which relies on one-hot binary encodings, the QUDO formulation directly represents the routing sequence using discrete-valued variables.

\paragraph{Decision variables}
We introduce discrete variables
\begin{equation}
v_j \in \{0,1,\dots,N\},
\quad j \in \{1,\dots,M\},
\end{equation}
where $M = N + V$ and $v_j = i$ indicates that node $i$ (customer or depot) occupies position $j$ in a cyclic sequence. Each variable $v_j$ is represented by a single qudit of local dimension $d = N+1$.

\paragraph{Objective function}
The total routing cost is given by
\begin{equation}
H_{\mathrm{cost}}
=
\sum_{j=1}^{M}
D_{v_j,\,v_{(j+1)\bmod M}},
\end{equation}
which directly encodes the distance between consecutive nodes in the cyclic sequence, including depot to customer and customer to depot transitions.

\paragraph{Constraints}
Feasibility is enforced using collision-based penalty terms. First, each customer must appear exactly once in the sequence. This is achieved by penalizing repeated occurrences of the same customer:
\begin{equation}
H_{\mathrm{city}}
=
A
\sum_{i=1}^{N}
\sum_{1 \le p < q \le M}
\delta(v_p = i)\,\delta(v_q = i),
\end{equation}
where $\delta(\cdot)$ denotes the Kronecker delta function. This term prevents two vehicles from visiting the same customer and eliminates the need for explicit one-hot constraints.

Second, the number of vehicles is enforced by requiring the depot to appear exactly $V$ times in the sequence:
\begin{equation}
H_{\mathrm{depot}}
=
B\left(
\sum_{j=1}^{M}
\delta(v_j = 0)
-
V
\right)^2.
\end{equation}
Each occurrence of the depot acts as a route separator, so that the subsequences of customers between consecutive depot positions define the individual vehicle routes.

\paragraph{QUDO Hamiltonian}
The complete QUDO energy function is
\begin{equation}
H_{\mathrm{VRP}}^{\mathrm{QUDO}}
=
H_{\mathrm{cost}}
+
H_{\mathrm{city}}
+
H_{\mathrm{depot}},
\end{equation}
with penalty weights chosen such that $A \gg B > 0$.

\section*{Appendix C}
\subsection*{Multi Depot VRP}
\subsubsection*{QUBO Formulation}
In MDVRP, we consider $D$ depots and $N$ customers served by a total of $V$ vehicles, where each vehicle starts and ends at a designated depot. Let $V_d$ denote the number of vehicles assigned to depot $d$, with $\sum_{d=1}^{D} V_d = V$. The problem is encoded as a QUBO model by extending the permutation-based formulation used for the traveling salesman problem.

We introduce binary decision variables
\begin{equation}
X_{i,j} \in \{0,1\},
\quad i \in \{0,1,\dots,N+D-1\}, \quad j \in \{1,\dots,M\},
\end{equation}
where $M = N + V$ denotes the length of a cyclic sequence. The index $i$ labels both customers and depots, with depot nodes allowed to appear multiple times, and $X_{i,j}=1$ indicates that node $i$ occupies position $j$ in the sequence.

\paragraph{Objective Function}
The routing cost is expressed as
\begin{equation}
H_{\mathrm{cost}} =
\sum_{j=1}^{M}
\sum_{i=0}^{N+D-1}
\sum_{k=0}^{N+D-1}
D_{i,k}\,
X_{i,j}\,
X_{k,(j+1)\bmod M},
\end{equation}
which accounts for all customer--customer, depot--customer, and customer--depot transitions along the cyclic tour.

\paragraph{Feasibility Constraints}
Feasibility is enforced through quadratic penalty terms. First, the position constraint
\begin{equation}
H_{\mathrm{pos}} =
A\sum_{j=1}^{M}
\left(\sum_{i=0}^{N+D-1} X_{i,j} - 1\right)^2
\end{equation}
ensures that exactly one node occupies each position in the sequence.

Second, customer uniqueness is enforced by
\begin{equation}
H_{\mathrm{cust}} =
A\sum_{i \in \mathcal{C}}
\left(\sum_{j=1}^{M} X_{i,j} - 1\right)^2,
\end{equation}
where $\mathcal{C}$ denotes the set of customer indices, guaranteeing that each customer is visited exactly once.

Third, depot multiplicity constraints ensure that each depot $d$ appears exactly $V_d$ times in the sequence:
\begin{equation}
H_{\mathrm{dep}} =
B\sum_{d \in \mathcal{D}}
\left(\sum_{j=1}^{M} X_{d,j} - V_d\right)^2,
\end{equation}
where $\mathcal{D}$ denotes the set of depot indices. Each occurrence of a depot acts as a route separator and implicitly defines a vehicle route.

Finally, consecutive depot placements are forbidden using the adjacency penalty
\begin{equation}
H_{\mathrm{adj}} =
C\sum_{j=1}^{M}
\sum_{d_1,d_2 \in \mathcal{D}}
X_{d_1,j}\,
X_{d_2,(j+1)\bmod M},
\end{equation}
which prevents depot--depot transitions and ensures that each vehicle serves at least one customer.

\paragraph{QUBO Hamiltonian}
The complete QUBO Hamiltonian for the multi-depot vehicle routing problem is given by
\begin{equation}
H_{\mathrm{MDVRP}}^{\mathrm{QUBO}} =
H_{\mathrm{cost}} +
H_{\mathrm{pos}} +
H_{\mathrm{cust}} +
H_{\mathrm{dep}} +
H_{\mathrm{adj}},
\end{equation}
with penalty weights chosen such that $A \gg B \gg C > 0$. Although this formulation exactly encodes the MDVRP, it requires $(N+D)(N+V)$ binary variables and induces dense quadratic couplings, leading to rapid growth in problem size and limiting scalability on near-term quantum hardware.

\subsubsection*{QUDO formulation}

We now formulate MDVRP as a QUDO problem. In contrast to the binary one-hot encoding used in QUBO, the QUDO formulation represents the routing sequence directly using discrete variables, leading to a more compact encoding and substantially fewer constraints.

\paragraph{Decision Variables}
Let $M = N + V$ denote the length of the routing sequence. We introduce discrete variables
\begin{equation}
v_j \in \{0,1,\dots,N+D-1\},
\quad j = 1,\dots,M,
\end{equation}
where $v_j=i$ indicates that node $i$ (either a customer or a depot) occupies position $j$ in the cyclic sequence. Customer labels appear exactly once, while depot labels may appear multiple times and act as route separators.

\paragraph{Objective Function}
The routing cost is expressed compactly as
\begin{equation}
H_{\mathrm{cost}} =
\sum_{j=1}^{M}
D_{v_j,\,v_{(j+1)\bmod M}},
\end{equation}
which accounts for all transitions between consecutive nodes in the cyclic tour, including customer--customer, depot--customer, and customer--depot edges.

\paragraph{Customer Uniqueness Constraint}
Each customer must be visited exactly once. This is enforced using collision-based penalties rather than one-hot constraints:
\begin{equation}
H_{\mathrm{cust}} =
A \sum_{i \in \mathcal{C}}
\sum_{1 \le p < q \le M}
\delta(v_p=i)\,\delta(v_q=i),
\end{equation}
where $\mathcal{C}$ denotes the set of customer indices. Any repeated appearance of the same customer incurs an energy penalty, thereby preventing multiple vehicles from visiting the same customer.

\paragraph{Depot Multiplicity Constraint}
To enforce the number of vehicles assigned to each depot, we require depot $d$ to appear exactly $V_d$ times in the sequence:
\begin{equation}
H_{\mathrm{dep}} =
B \sum_{d \in \mathcal{D}}
\left(
\sum_{j=1}^{M} \delta(v_j=d) - V_d
\right)^2,
\end{equation}
where $\mathcal{D}$ denotes the set of depot indices. Each occurrence of a depot acts as a route separator and implicitly defines one vehicle route.

\paragraph{Depot Adjacency Constraint}
To ensure that each vehicle services at least one customer, consecutive depot placements are forbidden using
\begin{equation}
H_{\mathrm{adj}} =
C \sum_{j=1}^{M}
\sum_{d_1,d_2 \in \mathcal{D}}
\delta(v_j=d_1)\,\delta(v_{(j+1)\bmod M}=d_2),
\end{equation}
which penalizes depot--depot transitions in the cyclic sequence.

\paragraph{QUDO Hamiltonian}
The complete QUDO Hamiltonian for the multi-depot vehicle routing problem is given by
\begin{equation}
H_{\mathrm{MDVRP}}^{\mathrm{QUDO}} =
H_{\mathrm{cost}} +
H_{\mathrm{cust}} +
H_{\mathrm{dep}} +
H_{\mathrm{adj}},
\end{equation}
with penalty weights chosen such that $A \gg B \gg C > 0$. This formulation encodes all structural constraints of the MDVRP without introducing auxiliary binary variables or position-wise one-hot constraints. As a result, the problem requires only $M=N+V$ qudits of local dimension $N+D$, offering an exponential reduction in the number of degrees of freedom compared to the corresponding QUBO formulation.

\section*{Appendix D}
\subsection*{Max-K-Cut}
\subsubsection*{QUBO formulation}
\paragraph{Decision variables.}
The standard QUBO encoding introduces binary one-hot variables
\begin{equation}
x_{i,k} \in \{0,1\},
\quad i=1,\dots,N,\; k=1,\dots,K,
\end{equation}
where $x_{i,k}=1$ indicates that vertex $i$ is assigned to partition $k$.

\paragraph{Objective function.}
An edge $(i,j)\in E$ contributes one unit to the cut if its endpoints are assigned
to different partitions. This condition can be written as
\begin{equation}
1 - \sum_{k=1}^{K} x_{i,k}x_{j,k}.
\end{equation}
The total cut size is therefore
\begin{equation}
\sum_{(i,j)\in E}
\left(1 - \sum_{k=1}^{K} x_{i,k}x_{j,k}\right).
\end{equation}

\paragraph{QUBO Hamiltonian}
The one-hot constraints are enforced via quadratic penalties. The full QUBO
Hamiltonian is
\begin{align}
H_{\mathrm{Max\text{-}K\text{-}Cut}}^{\mathrm{QUBO}}
&=
-\sum_{(i,j)\in E}
\left(1 - \sum_{k=1}^{K} x_{i,k}x_{j,k}\right)
\nonumber\\
&\quad
+ A \sum_{i=1}^{N}
\left(\sum_{k=1}^{K} x_{i,k} - 1\right)^2 ,
\end{align}
where $A>0$ is a penalty parameter chosen sufficiently large to ensure that
each vertex is assigned to exactly one partition. This formulation requires
$NK$ binary variables and introduces dense quadratic couplings through the
constraint terms.

\subsubsection*{QUDO Formulation}

\paragraph{Decision variables.}
In the QUDO formulation, each vertex is represented by a single $K$-ary variable
\begin{equation}
s_i \in \{1,\dots,K\}, \quad i=1,\dots,N,
\end{equation}
where $s_i$ denotes the partition label of vertex $i$.

\paragraph{QUDO Hamiltonian}
An edge $(i,j)\in E$ contributes to the cut if $s_i \neq s_j$. Using the Kronecker
delta $\delta(\cdot)$, the QUDO Hamiltonian for the unweighted Max-$K$-Cut problem
is
\begin{equation}
H_{\mathrm{Max\text{-}K\text{-}Cut}}^{\mathrm{QUDO}}
=
-\sum_{(i,j)\in E}
\left(1 - \delta(s_i=s_j)\right).
\end{equation}

This formulation requires no additional constraints, as each variable $s_i$
naturally assumes exactly one of the $K$ partition labels.

\section*{Appendix E}
\subsection*{Graph Coloring}
\subsubsection*{QUBO Formulation}

\paragraph{Decision variables.}
The standard binary encoding introduces one-hot variables
\begin{equation}
x_{i,k}\in\{0,1\}, \quad i=1,\dots,N,\; k=1,\dots,K,
\end{equation}
where $x_{i,k}=1$ indicates that vertex $i$ is assigned color $k$.

\paragraph{Constraints.}
A valid coloring requires (i) each vertex has exactly one color and (ii) no edge has equal colors.
These are enforced by quadratic penalties:
\begin{align}
H_{\text{one-hot}}
&=
A \sum_{i=1}^{N}\left(\sum_{k=1}^{K}x_{i,k}-1\right)^2,\\
H_{\text{edge}}
&=
B \sum_{(i,j)\in E}\sum_{k=1}^{K} x_{i,k}x_{j,k},
\end{align}
where $A,B>0$ are penalty weights.

\paragraph{QUBO Hamiltonian.}
The resulting QUBO energy function is
\begin{equation}
H_{\text{GC}}^{\text{QUBO}}
=
H_{\text{one-hot}} + H_{\text{edge}}.
\end{equation}
When the instance is $K$-colorable and penalties are chosen sufficiently large, the minimum
energy is $0$, and any ground state corresponds to a proper coloring \citep{Lucas_2014}.

\subsubsection*{QUDO Formulation}

\paragraph{Decision variables.}
In the QUDO encoding, each vertex is represented by a single $K$-ary variable
\begin{equation}
s_i \in \{1,\dots,K\}, \quad i=1,\dots,N,
\end{equation}
where $s_i$ denotes the color assigned to vertex $i$.

\paragraph{QUDO Hamiltonian.}
Proper coloring requires $s_i\neq s_j$ for all $(i,j)\in E$. Using the Kronecker delta,
the QUDO energy may be written as
\begin{equation}
H_{\text{GC}}^{\text{QUDO}}
=
B \sum_{(i,j)\in E} \delta(s_i = s_j),
\end{equation}
which penalizes monochromatic edges. Because each $s_i$ always takes exactly one of the $K$
labels, no separate one-hot constraint is required. As in the QUBO case, when the instance
is $K$-colorable the minimum energy is $0$.

\section*{Appendix F}
\subsection*{Job Scheduling}
\subsubsection*{QUBO Formulation}

\paragraph{Decision variables.}
The standard QUBO encoding introduces binary one-hot variables
\begin{equation}
x_{i,j} \in \{0,1\},
\quad i,j = 1,\dots,N,
\end{equation}
where $x_{i,j} = 1$ indicates that job $i$ is scheduled at position $j$ in the processing sequence.

\paragraph{Objective function.}
The weighted completion time contribution of job $i$ scheduled at position $j$ is proportional to $w_i p_i j$. The total scheduling cost is therefore expressed as
\begin{equation}
H_{\mathrm{cost}}
=
\sum_{j=1}^{N}
\sum_{i=1}^{N}
w_i\, p_i\, j\, x_{i,j}.
\end{equation}

\paragraph{Constraint penalties.}
A feasible schedule must satisfy two permutation constraints:
\begin{enumerate}
\item each position is occupied by exactly one job,
\item each job appears exactly once in the sequence.
\end{enumerate}
These constraints are enforced using quadratic penalty terms, yielding the QUBO Hamiltonian
\paragraph{QUBO Hamiltonian}
\begin{align}
H_{\mathrm{sched}}^{\mathrm{QUBO}}
&=
H_{\mathrm{cost}}
+
A \sum_{j=1}^{N}
\left(\sum_{i=1}^{N} x_{i,j} - 1\right)^2
\nonumber\\
&\quad
+
A \sum_{i=1}^{N}
\left(\sum_{j=1}^{N} x_{i,j} - 1\right)^2 ,
\end{align}
where $A>0$ is a penalty parameter chosen sufficiently large to enforce feasibility. This formulation requires $N^2$ binary variables and introduces dense quadratic couplings arising from the permutation constraints \citep{Lucas_2014}.

\subsubsection*{QUDO Formulation}

\paragraph{Decision variables.}
In the QUDO formulation, the schedule is represented by a sequence of discrete variables
\begin{equation}
v_j \in \{1,\dots,N\},
\quad j = 1,\dots,N,
\end{equation}
where $v_j = i$ denotes that job $i$ is processed at position $j$.

\paragraph{Objective function.}
The total weighted completion time is directly expressed as
\begin{equation}
H_{\mathrm{cost}}
=
\sum_{j=1}^{N}
w_{v_j}\, p_{v_j}\, j .
\end{equation}

\paragraph{Collision penalties.}
Each job must appear exactly once in the schedule. This condition is enforced using collision-based penalties rather than one-hot constraints:
\begin{equation}
H_{\mathrm{job}}
=
A \sum_{1 \le p < q \le N}
\sum_{i=1}^{N}
\delta(v_p = i)\,\delta(v_q = i),
\end{equation}
where $\delta(\cdot)$ denotes the Kronecker delta. This term penalizes repeated scheduling of the same job at multiple positions.

\paragraph{QUDO Hamiltonian}
\begin{equation}
H_{\mathrm{sched}}^{\mathrm{QUDO}}
=
H_{\mathrm{cost}} + H_{\mathrm{job}} .
\end{equation}

In contrast to the QUBO formulation, the QUDO encoding requires only $N$ discrete variables of local dimension $N$, with feasibility enforced directly through pairwise collision penalties.

\newpage
\bibliographystyle{unsrtnat} 
\bibliography{bibliography}

\end{document}